\PassOptionsToPackage{pdfpagelabels=false}{hyperref} 
\documentclass[fleqn,letters,onecolumn]{mnras}

\usepackage{newtxtext,newtxmath,pdflscape}

\usepackage[T1]{fontenc}
\usepackage{ae,aecompl}

\usepackage{graphicx}	
\usepackage{amsmath}	
\usepackage{amssymb}	

\def\etal   {{\it et al.\ }}             
\def\msol   {\hbox{$\hbox{M}_\odot$}}

\def\etal   {{\it et al.}}                     

\def\pdeg           {$.\kern-.25em ^{^\circ}$}
\def\degree{\ifmmode{^\circ} \else{$^\circ$}\fi}
\def\ut #1 #2 { \, \textrm{#1}^{#2}} 
\def\u #1 { \, \textrm{#1}}          

\def\le     {$\leq$}                            
\def\msol   {\hbox{$M_\odot$}}                  
\def\v      {\hbox{\it V}}                      

\def\etal   {{\it et al.}}                     
\def\la{\lower.4ex\hbox{$\;\buildrel <\over{\scriptstyle\sim}\;$}}
\def\ga{\lower.4ex\hbox{$\;\buildrel >\over{\scriptstyle\sim}\;$}}

\newcommand{\sgra}{Sgr~A$\textrm{*}$}

\newcommand   {\arcs}  {\mbox{$^{\prime\prime}$}}

\renewcommand {\ga}    {\mbox{\rlap{\hbox{\lower5pt\hbox{$\sim$}}}\hbox{$>$}}}
\renewcommand {\la}    {\mbox{\rlap{\hbox{\lower5pt\hbox{$\sim$}}}\hbox{$<$}}}

\title[Sgr A*]{ALMA and VLA observations of 
       emission  from the environment of Sgr A*}

\author[F. Yusef-Zadeh]{
F. Yusef-Zadeh,$^{1}$\thanks{E-mail: zadeh@northwestern.edu (FYZ)}
R. Sch\"odel$^2$
M. Wardle$^3$
H. Bushouse$^4$
W. Cotton$^5$
\newauthor
M. J. Royster$^1$
D. Kunneriath$^2$ 
D. A. Roberts$^1$ 
E. Gallego-Cano$^2$
\\
$^{1}$Department of Physics and Astronomy
Northwestern University, Evanston, IL 60208, US\\
$^{2}$Instituto de Astfisica de Andalucia (CSIC),
Glorieta de la Astronomia S/N, 18008 Granada, Spain\\
$^{3}$Department of Physics and Astronomy and
Research Centre for Astronomy, Astrophysics\\
\& Astrophotonics, Macquarie University, Sydney NSW 2109, Australia\\
$^4$Space Telescope Science Institute, Baltimore, MD 21218\\
$^5$National Radio Astronomy Observatory,  Charlottesville, VA 22903
}

\pubyear{2015}

\PassOptionsToPackage{pdfpagelabels=false}{hyperref} 
\begin{document}
\label{firstpage}
\pagerange{\pageref{firstpage}--\pageref{lastpage}}
\maketitle

\begin{abstract}
We present 44 and 226 GHz observations of the Galactic center within 20$''$ of Sgr A*. Millimeter 
continuum emission at 226 GHz is detected from eight stars that have previously been identified at
near-IR and radio wavelengths. 
We also detect a 5.8 mJy source at 226 GHz
coincident with the magnetar SGR~J1745-29 located 2.39$''$ SE of Sgr A* and identify a new
2.5$''\times1.5''$ halo of mm emission centered on Sgr A*. The X-ray emission from this halo has
been detected previously and is interpreted in terms of a radiatively inefficient accretion flow.
The mm halo surrounds an EW linear feature which appears to arise from Sgr A* and coincides with
the diffuse X-ray emission and a minimum in the near-IR extinction. We argue that the millimeter
emission is produced by synchrotron emission from relativistic electrons in equipartition with a
$\sim 1.5$\,mG magnetic field.  The origin of these is unclear but its coexistence with hot gas
supports scenarios in which the gas is produced by the interaction of winds either from the fast
moving S-stars, the photo-evaporation of low-mass YSO disks or by a jet-driven outflow from Sgr A*.
The spatial anti-correlation of the X-ray, radio and mm emission from the halo and the low near-IR
extinction provides compelling evidence for an outflow 
sweeping up the interstellar material, creating a dust cavity within
2$''$ of Sgr A*. Finally, the radio and mm counterparts to eight near-IR identified stars within
$\sim$10\arcs\ of \sgra\ provide accurate astrometry to determine the positional shift between the
peak emission at 44 and 226 GHz. 



\end{abstract}

\begin{keywords}
accretion, accretion disk -- black hole physics --  Galaxy: centre
\end{keywords}



\section{Introduction}

The nuclear region of our Galaxy coincides with a stellar cluster consisting of an evolved
population and a young population of OB and WR stars (Paumard \etal\ 2006; Lu \etal\ 2009) centered
on the 4$\times10^6$ \msol\ black hole \sgra\ (Reid and Brunthaler 2004; Genzel et al. 2010;
Sch\"odel et al. 2014; Boehle \etal\ 2016; Gillsessen \etal\ 2016). Until recently, the young
massive stars within 0.5 pc of Sgr A* could only be identified and studied employing adaptive
optics in the near-IR. Recent high-resolution radio continuum observations detected 318 compact
radio sources within the inner 30\arcs\ of \sgra. The comparison of radio and near-IR data indicate
that at least 45 of the compact radio sources coincide with known sources, many of which are
massive stars, identified at $K_s$ and $L'$ bands in the near-IR (Yusef-Zadeh et al. 2016).

Thermal radio emission from a mass-losing star arises from spherically-symmetric, wind of
fully--ionized gas expanding at its terminal velocity 
(e.g., Panagia \& Felli 1975). 
Stellar
thermal emission at radio wavelengths could also arise from the photospheres of evolved stars in
the nuclear cluster. One way to distinguish between these two different populations is to determine
their radio spectral index $\alpha$ where the flux density S$_\nu\propto\nu^{-\alpha}$.
Photospheric emission has an inverted spectrum with $\alpha\sim2$ whereas ionized stellar winds
from young massive stars typically have a spectrum with $\alpha\sim0.6$ (Panagia 1973).  Previous
detection of radio emission from near-IR identified stars in the inner 10$''$ of Sgr A*
(Yusef-Zadeh et al. 2015a) had a limited frequency coverage between 34 and 44 GHz, thus, it was not
possible to accurately determine the spectral index.  The present 226 GHz and 44 GHz observations
present an opportunity to remedy this by determining the spectrum of radio emission and
distinguishing between members of the evolved cluster and massive stars in the young cluster.

To measure orbits of stars around Sgr A*, it is necessary to tie radio and near-IR data into
astrometric reference frames. The positions of radio stars can provide precise astrometry relative
to Sgr A*. The detection of stars at millimeter (mm) wavelengths opens a new window for astrometric
calibration as well as examining if there is a shift in the peak position of Sgr A* between radio
and mm wavelengths as might be expected, for example if mm emission from a jet was present (Markoff
et al. 2007)

Here we present simultaneous ALMA, VLA and VLT observations of the Galactic center and determine
the spectral index between 44 and 226 GHz of eight stellar sources identified at H (1.63 $\mu$m)
band and two nonthermal sources Sgr~A* and the magnetar SGR~J1745-29. In addition, we determine the
peak position of Sgr A* at 226~GHz by registering the accurate positions of near-IR identified
stars detected at 44 and 226 GHz and then search for any shifts in position of Sgr A* at 44 and 226
GHz.  Finally, the observations presented here show sub-structures associated with Sgr A* and a
network of narrow features at radio and mm. We detect a halo of mm emission from the inner 2$''$ of
Sgr A* which coincides with low near-IR extinction. The size of this halo is similar to that
detected at X-rays. We suggest that the outflow from the ionized winds of the S stars orbiting Sgr
A* are responsible for this emission. We report tentative detection of faint narrow fibrils of
radio and mm emission. We interpret these striking features as arising from the interaction of
radial outflows from the Galactic center and the atmospheres of mass-losing stars.
\section{Observations and Data Reduction}
\subsection{Radio and Millimeter Data}

The ALMA,   VLA\footnote{the Karl G. Jansky Very
Large Array (VLA) of the National Radio Astronomy Observatory is a facility of
the National Science Foundation, operated under a cooperative agreement by
Associated Universities, Inc.}, and VLT observations were carried out
as part of a
multi-wavelength observing campaign to monitor the flux variability of Sgr A*. This campaign was led by
Spitzer and Chandra,  and  radio and mm observations were obtained  as part of
the director's discretionary time given to us to join the observing campaign.
Titan was initially used in ALMA observations (project code 2015.A.00021.S.)
as the flux calibrator and NRAO 530
was observed periodically to correct for any changes in phase and amplitude
as a
function of time. The spectral windows were centered at roughly: 216.2 GHz, 218.0
GHz, 231.9 GHz, and 233.7 GHz. Editing and calibration of the data were
carried out with OBIT (Cotton 2008) before all the spectral windows were averaged prior to
constructing final images.  Observations were made on July 12 and 18, 2016 and the images were combined
after scaling  the variable flux of Sgr A* during the two epochs of observations.

We carried out VLA B-array observations (program 16A-419) in the Q (7mm) and Ka (9mm) bands on July
12 and 18, 2016 at 44 and 34 GHz, respectively.  We used the 3-bit system, which provided full
polarization in 4 basebands, each 2 GHz wide. Each subband was made up of 64 channels and channels
were 2 MHz wide.  We used 3C286 to calibrate the flux density scale and used 3C286 and J1733-1304
(aka NRAO530) to calibrate the bandpass and J1744-3116 to calibrate the complex gains. A phase and
amplitude self-calibration procedure was applied to all data using the bright radio source Sgr A*. We
used OBIT (Cotton 2008) and CASA to construct radio and mm images. The positions of radio stars are
determined with respect to the absolute position of Sgr A*. The Ka band data were compromised by bad
weather conditions, thus we do not present those observations.

\subsection{VLT Data}

NACO/VLT was used as part of a coordinated observing campaign, as
described above. Ten hours of observations were granted but the
observations suffered from bad seeing conditions, with the coherence
time being mostly $\tau_{0}\lesssim3$\,ms and the isoplanatic angle
$\theta_{0}\approx1"$\,arcsec.  Consequently, the Adaptive Optics (AO)
performance was mostly poor, with the exception of a few short
intervals of good conditions. As is standard for NACO AO observations
of the GC, the loop of the AO system was closed on the $K_{s}\approx7$
supergiant IRS\,7, located about $5.5''$ north of Sgr\,A*. For this
work we use the best data set, which was obtained on  2016 July 12 during UT 04:26 to
04:58. It consists of observations in the $H-$band with the S27
camera. The exposure time was set to 3\,s. We used NACO's cube-mode
and obtained 26 sets of 21 exposures each, amounting to a total
integration time of $1638$\,s. Data reduction was standard (sky
subtraction, flat-field correction, dead-pixel interpolation) and all
the reduced exposures were aligned via the centroid of the star
IRS16\,C and mean-combined. Stellar positions and fluxes were
extracted with the {\it StarFinder} package (Diolaiti et al. 2000).
We also present the extinction map of the inner $30''$ derived from
star counts in the near-IR, details of which are explained in
Sch{\"o}del et al. (2010). Foreground stars were removed prior to
creating the extinction map, which thus provides a good measure of the
column density of the clouds in the Galactic center.

\section{Results}

\subsection{A 2.5$''\times1.5''$ Emission Halo Centered on  Sgr A*}

Figure 1a shows the inner 17$''\times18''$ of Sgr A* at 225 GHz where the brightest portion of the
mini-spiral structure associated with the three-arms of the minispiral (Sgr A West) are detected. The
N and E arms and the bar to the south of Sgr A* are prominent. The region within a few arcseconds of
Sgr A*, outlined schematically by a dashed semi-circle, shows a number of new structures. One is a
diffuse halo structure with an elliptical appearance within 2.5$''\times1.5''$ of Sgr A* with mean
flux density of $\sim0.2-0.5$ mJy beam$^{-1}$.  The integrated flux is 81 mJy over an area of 9.7
square arcsecond. This diffuse structure has an X-ray counterpart (Wang et al. 2013). Figure 1b shows
a composite image of mm and X-ray emitting gas where we note the diffuse X-ray emission is coincident
with the mm halo centered on Sgr A*. The NS elongated X-ray structure G359.945--0.044 about $10''$ NW
of Sgr A* is a pulsar wind nebula candidate (Wang et al. 2013).  Figure 1c shows the composite image
of the extinction in the near-IR and the mm emission.  The extinction value in the Ks band
(2.17$\mu$m) ranges between 2.3 and 3.2 (Sch\"odel et al. 2010). We note the halo structure has the
lowest extinction $\sim2.4$ magnitudes compared to the high extinction of $\sim2.9$ magnitudes
surrounding it. Dark dashed lines to the south of Sgr A* outline the elliptical halo structure
bounded by the bar to the south of Sgr A*.  We also note the high extinction associated with the N
and E arms and the bar of the mini-spiral (Sch\"odel et al. 2010). The extinction map of the
mini-spiral shows clearly the complex nature of cold and dense gas that is externally photoionized by
the Galactic center radiation field to create the minispiral.

We note a bow-shock-like structure with an extent
of $\sim2''$ to the NE of Sgr A*. The typical surface brightness of this feature is about 0.2 mJy
beam$^{-1}$ at 226 GHz.  Finally, a  EW mm ridge of emission protrudes  from Sgr A*.
This ridge appears to extend further to
the west for several arcseconds before it merges with the continuum emission from the mini-spiral.

A close-up view of the inner 6$''\times4''$ of Sgr A* is shown
in Figure 2a where contours of mm emission are superimposed on  1.5-7 keV X-ray emission.
The extended X-ray  emission was identified by Wang et al. (2013).
This image shows  a
spatial correlation between mm and X-ray emission.
The diffuse X-ray and mm emission from the halo coincide with a region of low extinction.
The mm contours show an elliptically-shaped 2$.5''\times1.5''$ halo
of diffuse emission  similar to the X-ray morphology.
The anti-correlation of high emission  and low extinction suggest clearly that
the low extinction is caused by an outflow.
Figure 2b shows the close-up view of 226 GHz emission from Sgr A* and its vicinity.
The elongated  EW structure or  the linear ridge  appears to be associated with Sgr A*.
The  bow shock feature lies to  the eastern edge of the diffuse halo.  
The bow-shock may indicate where an outflow from Sgr A* interacts with the ISM in the immediate
vicinity of Sgr A*." 
We also note compact sources that coincide with stellar sources, as described below.

\subsection{Compact Stellar Sources  at  226 GHz}

Ten compact sources at 44 and 226 GHz are identified within the inner 10$''$ of \sgra. These sources,
labeled on Figures 2b and 3a,b, are used to astrometrically register the mm and near-IR images.  As
all radio, mm and near-IR data are taken at the same epoch, we identified radio and mm stars from the
first ALMA epoch 2016.54 by comparing all three images. ALMA observations detect eight near-IR
identified stellar sources at 226 GHz. These stars also have  counterparts at 44 GHz.
Tables 1 and 2 list Gaussian-fitted positions of 10 radio and mm sources distributed within the inner
10$''$ of Sgr A* at 44 and 226 GHz. 
Entries in the columns of  Tables 1 and 2 give the source name at 44
and 226 GHz, alternative names in the literature, the RA and Dec, the angular distance from Sgr A* in
increasing order, positional accuracy, the size of the source, the peak and integrated intensities.
Column 9 of Table 1 gives the spectral index of individual sources between 44 and 226 GHz. The last
column provides the comments on individual sources. The positional accuracy is determined from
quadrature sum of errors of the right ascension and declination values from 2D Gaussian fits without including 
absolute astrometric errors.
 Because the second epoch of
ALMA data had a higher spatial resolution, we compared this epoch with the first epoch of VLA
observation at 44 GHz to determine the spectral index of all sources except Sgr A*. The spectral
index of Sgr A* is estimated from the same epoch data sets. The positions and the sizes of radio
sources are determined from background-subtracted Gaussian fit to the individual radio sources.

We note the radio source associated with IRS 21 has a mm counterpart. This young stellar source
(Sanchez-Bermudez \etal\ 2014) is comprised of five radio components (Yusef-Zadeh \etal\ 2014) but
only one stellar source is identified at H-band.  The radio and infrared properties 
are similar to those of IRS 13N and IRS 13E suggesting that the radio emission arises from the disks
of massive YSO candidates in this cluster (Yusef-Zadeh et al. 2015b). We also detect mm emission from
the Galactic center magnetar, SGR~J1745-29 (Kennea et al.  2013; Shannon and Johnston 2013; Torne
\etal\ 2017). This source was in its quiescent phase before it was identified as an X-ray outburst
(Kennea et al. 2013). SGR J1745-29 is the closest known pulsar to Sgr A* located 2.4\arcs\ from
\sgra. The detection of a compact radio source was reported at $\alpha, \delta\, (J2000) = 17^h 45^m
40^s.16795\pm0.00002\, -29^{\circ} 00' 29''.74908\pm0.00064$ at 44.6 GHz on 2014 February 21
(Yusef-Zadeh \etal\, 2014). Our 44 GHz image shows a plume-like structure to the north of the
magnetar. This plume-like feature with an extent of $0.2''\times0.5''$ (width$\times$length)
 widens to the north with a peak flux density 0.8 mJy beam$^{-1}$. It does not have a counterpart at
3.8$\mu$m (Eckart \etal\ 2013) and has no obvious counterpart at 226 GHz. We do not have sufficient
data to determine its spectral index. Future polarization and spectral studies of this feature would
determine its nature.

The spectral index of Sgr A* $\alpha=0.56$ listed in Table 1 is steeper than 
 previously determined from  snapshot measurements (An 
et al. 2005). The magnetar has a relatively flat spectrum with $\alpha=-0.21$. The mm emission from the 
magnetar could be due to the combination of pulsed and diffuse shocked emission produced by the 
interaction of the pulsar outburst with the ISM (Yusef-Zadeh et al. 2016). The remaining eight sources 
are stellar, six of which have spectra consistent with ionized winds. Sources 7 and 8,  which coincide with 
IRS 3 and IRS 7SW, respectively, have a steep optically thick spectrum  
consistent with  photospheric radio emission. Alternative, these steep spectrum sources could also 
be generated 
by ionized winds of massive stars with
or wind sources with 
a varying density gradient and geometry (Panagia \& Felli 1975). 
IRS 3 is the brightest and most
extended 3.8$\mu$m Galactic center (Pott \etal\ 2008) stellar source.  The asymmetric shape of the
IRS 3 envelope may reflect tidal distortion by Sgr A* (Yusef-Zadeh et al. 2017).

\subsection{Search for Radio and mm Emission from the S stars}

There is a cluster of B dwarf stars associated with the S-star cluster within 1$''$ of Sgr A*
(Gillessen et al. 2009, 2016; Yelda et al. 2010, 2014). The detection of S stars in the radio and mm
has been challenging because of the bright and variable source Sgr A* and its 
frequency dependent  angular size  due to interstellar scattering.  To search for
radio emission from stars within 1$''$ of Sgr A*, we first calculated the positions of the S stars at
the epoch of the mm observation on 2016, July 16 by using orbital parameters derived from near-IR
observations (Gillessen et al. 2016). Table 3 gives the positions of the S cluster members offset
from Sgr A* in R.A. and Declination and their corresponding positional uncertainties at the epoch of
2016.54. The expected positions are indicated as crosses in three images, taken within a few days of
each other, a 1.6$\mu$m H-band, 226 GHz and 44 GHz image, as shown in Figures 4a-c, respectively. The
S stars are superimposed after astrometric corrections have been applied to the near-IR and mm
images.  The near-IR stellar sources coincide well with the predicted positions in Figure 4a. Two
bright stellar sources S96 and S97 in Figure 4b coincide with IRS 16NE and IRS 16SE, respectively.
There are also coincidences between some S stars, such as S83, and mm peaks in Figure 4b. However,
there is extended mm emission, and it is not possible to localize mm counterparts to stellar sources,
mainly because of the confusing compact and extended sources.  Remarkably, we note a number of 44 GHz
sources throughout the inner 3$''$ of Sgr A*, as shown in Figure 4c. In particular,
 S83 and S33 may have radio counterparts within the positional errors at 44 GHz. Similarly, S5
and S14 coincide with peak radio emission at a level of 0.15 mJy beam$^{-1}$.  Proper motion
measurements in the radio are needed to establish that they are indeed counterparts to S stars.

We note a number of radio and mm peaks without stellar counterparts in the H band.  Radio sources
could have counterparts in the L$'$ band where a number of dusty sources or the so-called dusty S
cluster objects (DSO/G2) have been detected (Eckart et al. 2014).  However, we have neither the
proper motion of DSO sources nor an L$'$ band image taken in the same epoch as the data presented
here to confirm 3.8$\mu$m counterparts to radio peaks. A member of this class of objects detected in
the L$'$ band is G2 that was the subject of intense observational campaigns as it passed extremely
close to the central black hole (Gillessen et al. 2012; Witzel et al. 2014; Pfuhl et al. 2015;
Valencia-Schneider et al. 2015). Given the low spatial resolution of our radio and mm observations,
we could not search for radio counterparts to G2. The expected position of the G2 cloud is -85 and 80
mas from Sgr A* (Gillessen, private communication) which is smaller than 
the 222$\times128$ mas synthesized beam 
at 44 GHz.

The compact radio and mm sources detected within 2$''$ of Sgr A* with no stellar counterparts could be
massive young stellar objects (YSOs), similar to radio counterparts of several members of the IRS 13N
cluster (Eckart et al. 2013; Yusef-Zadeh et al. 2014, 2015b). In this interpretation, ionized gas is
being photo-evaporated from the disks of YSOs by the UV radiation from young and massive stars
located between 1 and 10$''$ from Sgr A* (Yusef-Zadeh et al. 2014). Future proper motion and spectral
measurements of radio and mm sources are critical to determine their nature.

\subsection{Dust Cavities near Sgr A*}

We also examined the physical relationship between the S stars and a dust cavity centered on Sgr A*
that is present in the extinction map of Sgr A West, as shown in Figure 1c (Sch\"odel et al. 2010).
Figure 4d shows the inner 4$''\times3.5''$ of Sgr A* where the largest concentration of S stars is
detected in the Sgr A* dust cavity. The dust cavity coincides with the elongated mm halo and excess
diffuse X-ray and radio emission, as discussed earlier (see Figures 2a,b and 4b,c). In addition, the
correlation of the dust cavity and a halo of enhanced emission at multiple wavelengths suggests an
outflow has destroyed or swept the cold and dense material away from Sgr A*.  This X-ray filled
cavity with minimum extinction provides the strongest evidence for an outflow, the origin of which is
discussed in the next section.

A larger view of the near-IR extinction map identifies regions of low and high columns of dust.
Figure 5a,b show the inner 13$''$ of Sgr A* delineating extinction clouds and 44 GHz radio continuum
emission, respectively. We note a cloud that lies along the northwestern extension of the E arm at
$\sim-5''$ and $\sim1''$ W and N of Sgr A*, respectively. The anticorrelation of this dust cavity and
a gap in the ionized gas strongly suggest that the dust cavity is associated with the ionized material. 

The extinction map shown in Figure 5a (Sch\"odel et al. 2010) reveals a second dust cavity $\sim3''$
to the NE of the Sgr A* (drawn as a circle on Fig. 5a). The elongation and the position angles of the
dust cavities are similar to the position angles of a tentative jet driven outflow from Sgr A* which
was reported by Yusef-Zadeh et al. (2012).  In addition, a number of resolved sources, X3, X7, F1,
F2, F3, P1, P4 and the Sgr A East tower show elongated structures with similar position angles in
radio and 3.8$\mu$m images (Muzic et al.  2007, 2010; Yusef-Zadeh et al. 2016). These elongated
emitting features combined with elongated dust cavities, centered on Sgr A* and to the NE of Sgr A*,
provide support for a common origin for the collimated outflow from Sgr A* at a position angle of
$\sim60^\circ$ (Sch\"odel \etal\ 2007).

Another piece of evidence suggestive of an outflow from Sgr A* is an elongated feature from Sgr A*
that curves to the SW for $\sim5''$ and terminates in the mini-cavity. Figure 6a,b show grayscale
images of the inner 4$''\times6''$ of Sgr A* and contours of 44 GHz emission superimposed on a H-band
image.  The outline of the elongated edge-brightened balloon-shaped feature with a 4$''\times1''$
extent, is drawn on Figure 6a. The northern arc-like structure of this balloon-shaped structure, as
shown in Wardle and Yusef-Zadeh (1992), coincides with a blob known as ``epsilon" (Yusef-Zadeh
\etal\, 1990) which has been detected in earlier low spatial resolutions at 15 GHz (Zhao \etal\,
1991). The new image shows clearly that the blobs are extended and continue to form a balloon in the
direction of the mini-cavity.  A balloon-shaped structure to the SW and a dust cavity to the NE of
Sgr A* suggest that these features are related. The proper motion measurement of the $\epsilon$ blob
shows high velocity ionized gas moving away from Sgr A* to the SW (Zhao \etal\, 2009). This
morphological and kinematic information suggests that the elongated features to the NE and SW are
consistent with the outflow interpretation from Sgr A*. Future high resolution proper motion,
polarization and spectral index measurements of the ridge will provide additional constraints on the
claim that this feature is physically associated with Sgr A*.

\subsection{Near-IR and Millimeter Astrometry}

Given the accurate position of Sgr A* at radio wavelength (Reid et al. 1999; Choate and Yusef-Zadeh
1999), we investigated if there were any positional shifts between the peak mm and radio emission
from Sgr A*. The detection of several stellar sources and the magnetar at both radio and mm
wavelengths in the same epoch provides a means of registering the Galactic center at radio and
near-IR frames. We calibrated the images astrometrically by using the common positions of sources
listed in Table 1. The positions in the near-IR images were measured via Gaussian fitting in AIPS.
The astrometry was solved with the IDL solve-astro routine from ASTROLIB. No distortion solution was
fitted, instead the linear terms were utilized, which involves slight shifts that exist between the 2
frames (VLT vs VLA). The VLT image was shifted in RA and Dec by 1.14 and 0.99 pixels with a pixel
size of 29.033 milliarcseconds (mas). The rms scatter was 0.40 and 0.71 amongst the sources. While
the VLA data should intrinsically have very good absolute pointing, the sources are so faint that the
centroids of each source are fairly uncertain.  The WCS coordinates of the VLT image were then
modified to adjust for the shifts. The RA/Dec coordinates of the common sources were then computed
which resulted in a better agreement to the VLA frame.

We included the uncertainty of the position of six stellar sources detected in both 226 (the first
epoch) and 44 GHz in the computation to register self-calibrated VLA and ALMA images.  The pointing
center of ALMA observations was set at $\alpha, \delta (J2000) = 17^h\ 45^m\ 40^s.040, -29^\circ\
00'\ 28''.2$.  After calibrating the data using CASA and applying self-calibration gains, the peak
position of Sgr A* is $\alpha, \delta\, (J2000) = 17^h\ 45^m\ 40^s.040004, -29^\circ\, 00'\
28''.1997$.  The computed shifts that the 226 GHz image needed to register radio and mm sources are
$-0.6524$ and $-5.821$ pixels with 29.033 mas pixel$^{-1}$. We obtained the astrometrically-corrected
position of Sgr A* at 226 GHz and determined that it is 3.37$\pm0.04$ mas to the east and 
11.03$\pm0.23$
mas to the north of the radio position, thus 
the centroid of Sgr A* at 226 GHz is shifted by $11.53\pm0.23$ mas NE of the radio position at 44 GHz. 
 This positional shift estimate assumes that the sources of
ionized winds are symmetrical at 44 and 226 GHz, thus there is no optical depth effect that can be
significant to explain the appearance of a shift in the position of Sgr A*.

Given that interstellar scattering is much smaller at 226 GHz than at 44 GHz, the origin of the
positional shift between radio and mm is not clear but it is likely that Sgr A* is contaminated by a
strong mm source that shifts the bright position of Sgr A*.  The source is resolved in Table 2 based
on our second observation on 2016, July 18. The deconvolved angular size from our first epoch
observation 2016, July 12 is 0.015$''\times0.013''$ (PA=122$\pm30^\circ$). If the linear ridge in
Figure 2b becomes brighter close to the peak of Sgr A* at 226 GHz, it might be responsible for the
shift. If so, this jet-like linear ridge arising from Sgr A* must have a hard or highly inverted
spectrum since there is no significant emission detected at lower frequencies. In fact, recent high
resolution 86 and 230 GHz observations of Sgr A* with milliarsecond resolutions show an asymmetric
source structure (Brinkerink et al. 2016; Fish et al. 2016). This secondary component has a position
angle PA$\sim90^\circ$ and is shifted 100 $\mu$as to the east of the main source.  This asymmetry
could be explained by interstellar scattering effects or intrinsic to the source (Brinkerink et al.
2016). 
The  positional shifts in  these and VLBA measurements 
with vastly different spatial
resolution suggests that they may be physically associated with each other and that the asymmetric
source structure of Sgr A* is likely to be an intrinsic jet-like source emitted in the east-west
direction before gets redirected to the northeast. 

\subsection{Radial Fibrils  of mm Emission}

One of the striking features we note in the mm images of the Galactic center are faint linear 
features with an extent ranging between 2 and 10$''$. The widths of these narrow features, which we
call fibrils, are unresolved spatially and their typical intensity is $\sim$50-100 $\mu$Jy
beam$^{-1}$ above the background.  Figure 7a points to seven fibrils that are tentatively detected
mainly to the NW and SW of Sgr A* at 226 GHz. Figures 7b with a slightly lower spatial resolution
shows the innermost region of Figure 7a where radial fibrils are mainly found in the direction away
from the Galactic center. Although weak linear features in complex radio images of Sgr A* can be
problematic, it is difficult to see how the linear features in ALMA images could be artifacts. This
is because of the geometry of the array which is has a spiral pattern unlike the linear configuration
of the VLA.  We also note that some of the fibrils appear to terminate at compact sources identified
as radio and mm mass-losing young star to the SW of Sgr A*. Figure 8a shows a blow-up image of the 
region in reverse color where a network of fibrils with strongest emission 
is detected. The lines $a$ to $d$  are drawn parallel to the PA of individual fibrils. The 
best example of these faint sources witha varying background is $b$. 
Fibrils $c$ and $d$ appear to arise from stellar windy sources AF and AFNW.  
We also compared the mean value of the region where fibrils are detected  in the region immediately to 
the east where there is no evidence of fibrils;  the comparison showed a six-time-increase  in the mean 
 flux density  of the region where fibrils are detected 
 but the rms value  was 0.1 mJy beam$^{-1}$  in both regions. 
Another indication of the fibrils, along the line $c$ in Figure 8a,  
can be seen in Figure 8b showing  a linear feature in the
direction away from AF/HH with an extent of 20$''$. In order to bring out the weak and extended tail
behind the AF star, the brightness of the mini-spiral is saturated. A schematic diagram of Figure 9
shows the faint fibrils as well as the dust and molecular layers (in black) associated with the
mini-spiral. Given these caveats, sensitive measurements are clearly needed to confirm the tentative
detection of fibrils described here.

\section{Discussion}

\subsection{Low Extinction Millimeter Halo}

$Chandra$ observations have characterized the X-ray emission surrounding Sgr A* as spatially extended
with a radius of $\sim1.5"$ (Baganoff et al. 2003; Wang et al. 2013; Rozanska et al. 2015). The X-ray
luminosity is interpreted in terms of a radiatively inefficient accretion flow (RIAF, e.g. Yuan et
al. 2004; Moscibrokzka et al. 2009). In this model, a fraction of the gaseous material accretes onto
Sgr~A* and the rest is driven as an outflow from Sgr A* (e.g., Quataert 2004; Shcherbakov and
Baganoff 2010; Wang et al. 2013). Alternatively, the diffuse X-ray emission associated with Sgr A* is
interpreted as an expanding hot wind produced by the mass-loss from B-type main sequence stars,
and/or the disks of photoevaporation of low mass young stellar objects (YSOs) at a rate $\sim
10^{-6}\msol\,\mathrm{yr}^{-1}$ (Yusef-Zadeh et al. 2016).  The new millimeter halo emission and a
dust cavity provide additional constraints on the origin of the gas in the inner 1$''$ of Sgr A*.

The millimeter halo is coincident with the X-ray emission around Sgr A*, which is dominated by 
bremsstrahlung arising from a medium with $n_e\sim 150\ut cm -3 $ and $T\sim 3\times10^7$\,K.  The
bremsstrahlung contribution at 230 GHz, about 0.2 $\mu$Jy, is negligible.  Thermal continuum from
dust can also be ruled out because of the halo's extinction deficit of 0.5 magnitudes at H-band
relative to its surroundings.  The millimeter emission could, however, be produced by synchrotron
emission from relativistic electrons in equipartition with a $\sim 1.5$\,mG magnetic field.  The
energy density of each of these components would then be $\sim10$\% of the thermal energy density of
the hot gas, so this is plausible.  The luminosity in the mm is $4\pi d^2\nu S_\nu \sim
1.4\times10^{33}\u erg \ut s -1 $, comparable to the X-ray luminosity, $L_X \sim 1\times10^{33}\u erg
\ut s -1 $, implying that synchrotron cooling is marginally the dominant cooling mechanism for the
gas.  The synchrotron cooling time is $\sim1000$\,yr, cf. the hot gas cooling time $\sim10^{5}$\,yr,
so this requires electron acceleration  on this time scale.

The coexistence of synchrotron emission with the hot gas supports a scenario in which the gas is
produced by the interaction of winds either from the S-stars or by the photo-evaporation of low-mass
YSO disks (Yusef-Zadeh at al 2016).  In this picture the high relative speed of the orbital motion of
the sources means that the gas is shocked to keV temperatures even though the outflow velocity from
the sources is low.  These shocks would also accelerate relativistic electrons.  In steady state, the
rate of conversion of kinetic energy to heat and relativistic electrons in shocks should equate to
the X-ray and mm luminosities, respectively, therefore they should be approximately equal.

An alternative scenario is that outflow from the vicinity of Sgr A* has created an X-ray/mm bubble in
a denser medium. The extinction deficit $A_K \approx 0.5$ associated with the bubble is equivalent to
a ``missing'' column density $\sim10^{22}\ut cm -2 $, or a number density $n_H\sim3\times10^4\, \ut
cm -3 $.  The current pressure inside the bubble would drive expansion at a speed of
$\sim40$\,km\,s$^{-1}$ into such a medium, yielding an expansion time scale of $\sim1300$\,yr.

One possible origin of the outflow is from mass-losing evolved and/or young stars. Because of the
inverted spectrum of mass-losing stars the emission is stronger at at higher frequencies. Assuming a
typical flux density $\sim1$ mJy at 226 GHz, a total of 80 stars are needed to account for the
diffuse emission. So, in this picture, the diffuse millimeter emission should resolve into individual
stars with stellar winds. Because of the high orbital motion of stars and higher mass loss rates,
$\sim 10^{-7}$--$10^{-6}$\,\msol\,yr$^{-1}$ from recent modeling (Offner \& Arce 2015), the X-ray gas
is supplied by shocked winds.  Similar to mm emission, X-ray emission should be resolved into
individual stellar sources. One possibility is enhanced X-ray emission at about 1.06$''$ from the
peak X-ray emission from Sgr A*. This emission appears to coincide with IRS 16C.

In summary, we have presented a variety of structures within 30$''$ of Sg A* (see the diagram in
Figure 9) using data taken with the VLA and ALMA.  On the smallest scale, we detect 226 GHz emission
from a $2.5''\times1.5''$ halo that appears to coincide with a dust cavity and diffuse X-ray gas
centered on Sgr A*. This mm emission coincides with diffuse X-ray emission centered on Sgr A*. We
argued that the mm emission is due to synchrotron, generated either from fast-moving orbiting stars
or protostars or from the activity associated with Sgr A*. This implies an outflow that produced the
mm and X-ray emission and destroyed dust grains. On a scale of 5$''$ from Sgr A*, we detect elongated
balloon-shaped structure and a dust cavity that are roughly in the direction where a number of
head-tail radio sources are found in previous measurements. These morphological details can be
described by a collimated outflow from Sgr A* at a position angle of 60$^\circ$.  We also detected mm
emission from ionized winds of massive stars orbiting Sgr A* and determined their spectral indices.
Lastly, we found a discrepancy in the peak position of Sgr A* between radio and 226 GHz. Future high
frequency ALMA observations should be able to place a better constraint on frequency-depended
position of Sgr A* and to confirm tentative detection of a network of faint fibrils distributed
throughout the inner 15$''$ of the Galactic center.

Acknowledgments:
We thank the referee for excellent comments. 
This work is partially supported by the grant AST-0807400 from the NSF and the European Research
Council under the European Union's Seventh Framework Program (FP/2007-2013). The research leading to
these results has received funding from the European Research Council under the European Union's
Seventh Framework Program (FP7/2007-2013) / ERC grant agreement number 614922] and by an Outside
Studies Program Fellowship awarded to Macquarie University. The National Radio Astronomy Observatory
is a facility of the National Science Foundation, operated under a cooperative agreement by
Associated Universities, Inc.






\vfill\eject\

\begin{figure}
\center
\includegraphics[scale=0.35,angle=0]{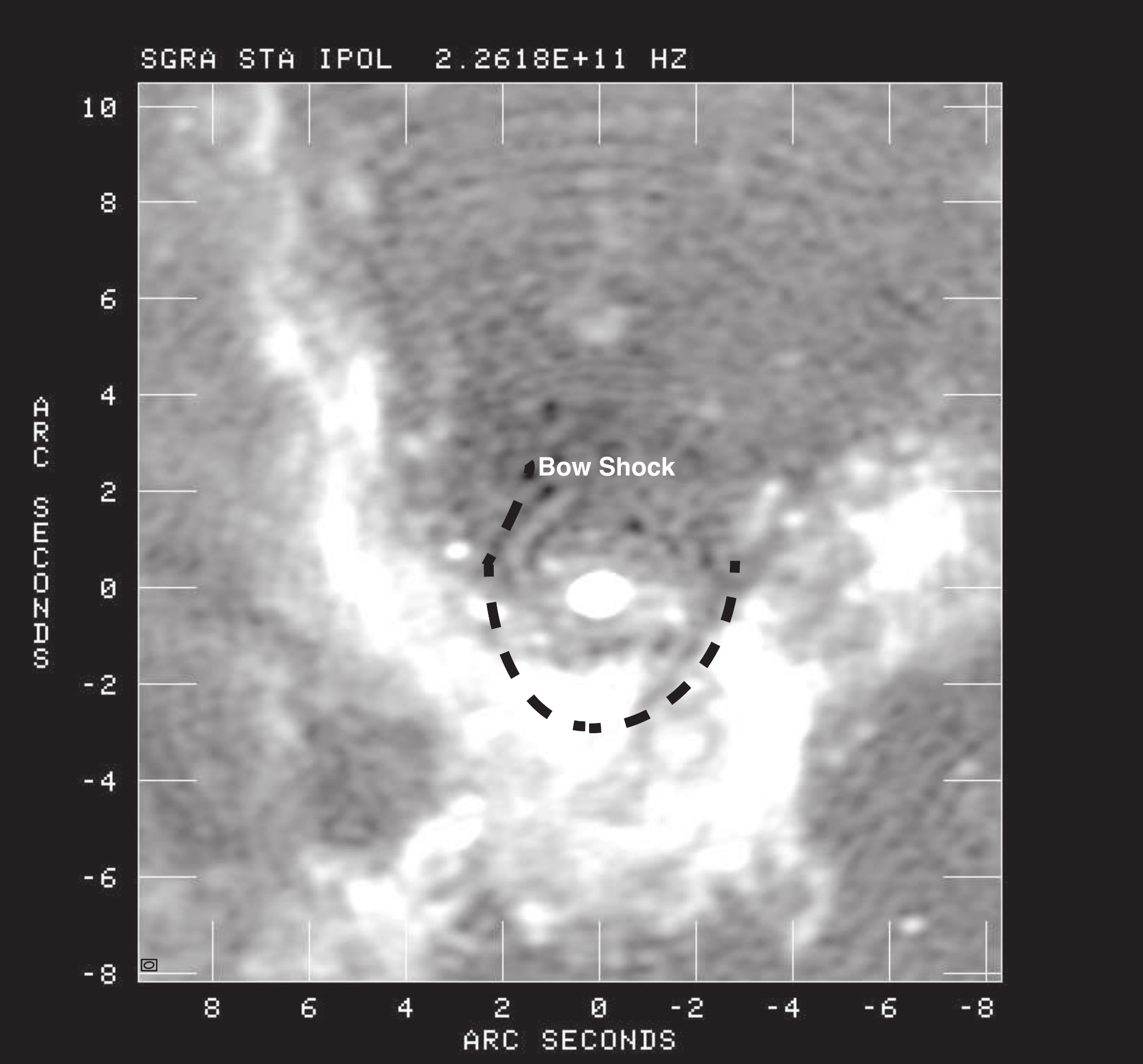}\\
\includegraphics[scale=0.35,angle=0]{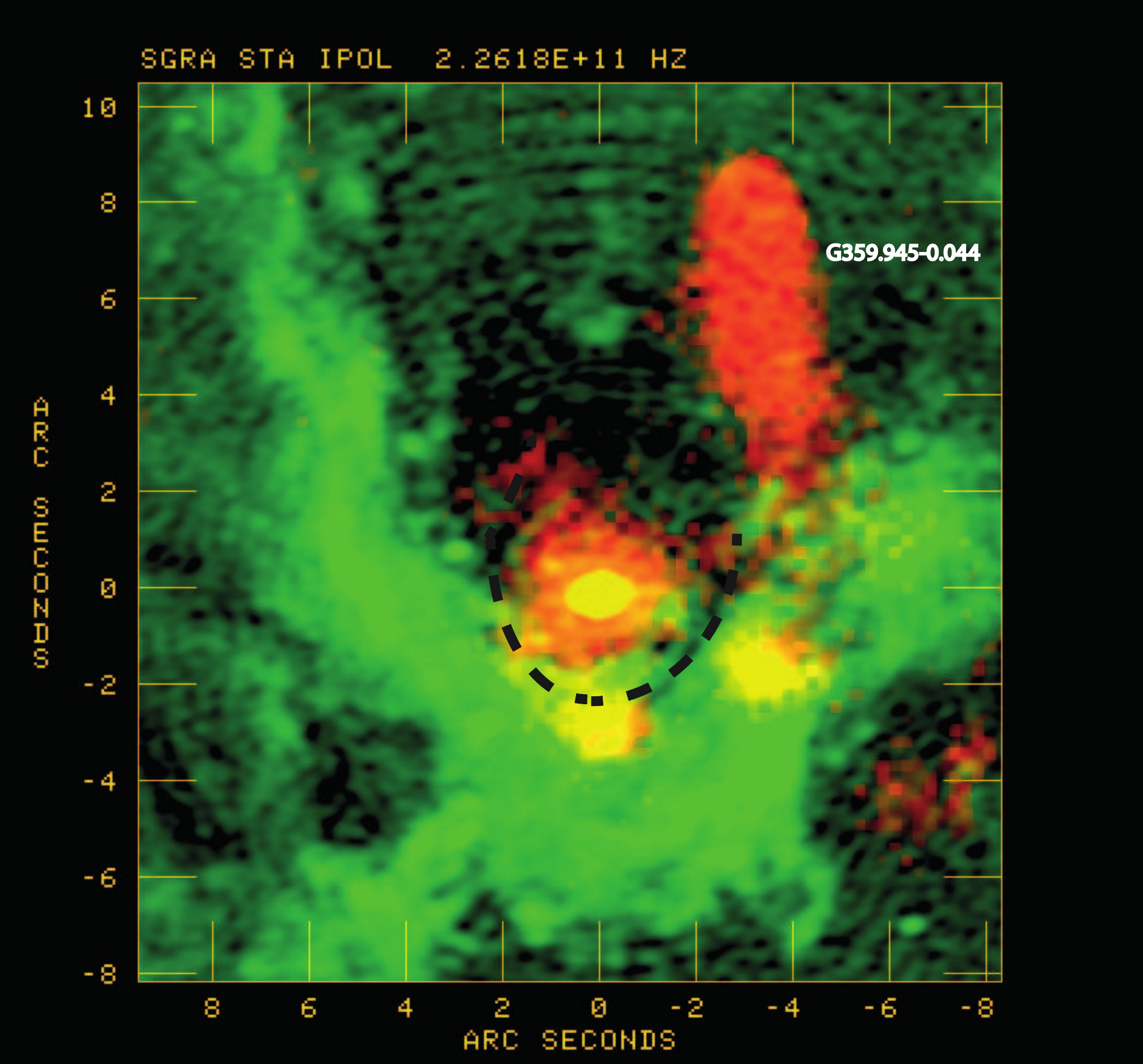}\\
\includegraphics[scale=0.35,angle=0]{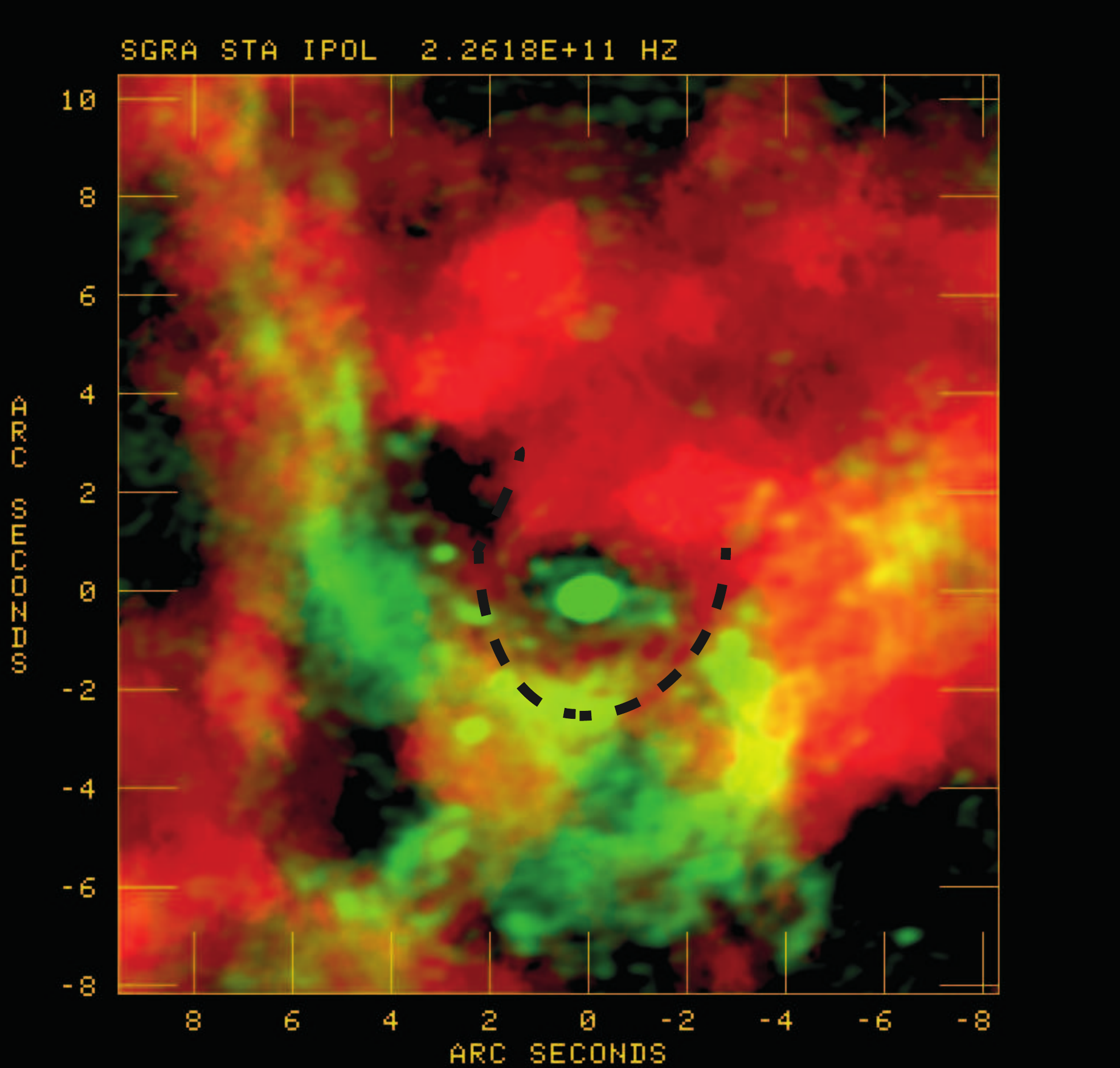}
\caption{
{\it (a)}
A 226 GHz image of the mini-spiral with spatial resolution of 0.38$''\times0.27''$ and 
PA=$-79^\circ$. This image is based on combining both observations taken in two epochs on 2016, July 
12 and 18. The peak flux is 3.47 Jy beam$^{-1}$. (The grayscale range $-6.7\times10^{-4} - 
3\times10^{-3}$ Jy beam$^{-1}$.) {\it (b)} Similar to (a) except that an X-ray 1.5-7 keV image in red 
taken with Chandra (D. Haggard, private communication) is superimposed on a 226 GHz image in green.  
(The grayscale range 0 to 10$^4$ counts.)
$-6\times10^{-4} - 3\times10^{-3}$ Jy beam$^{-1}$.)
{\it (c)} Similar to (a) except the Ks extinction image (Sch\"odel et al. 2010) in red 
is superimposed on a 226 GHz image in green.  The range of extinction value is between 
2.35 and 3.20 magnitudes.
(The grayscale range $-6.7\times10^{-4} - 1\times10^{-2}$ Jy beam$^{-1}$.)
}
\end{figure}

\begin{figure}
\center
\includegraphics[scale=0.6,angle=0]{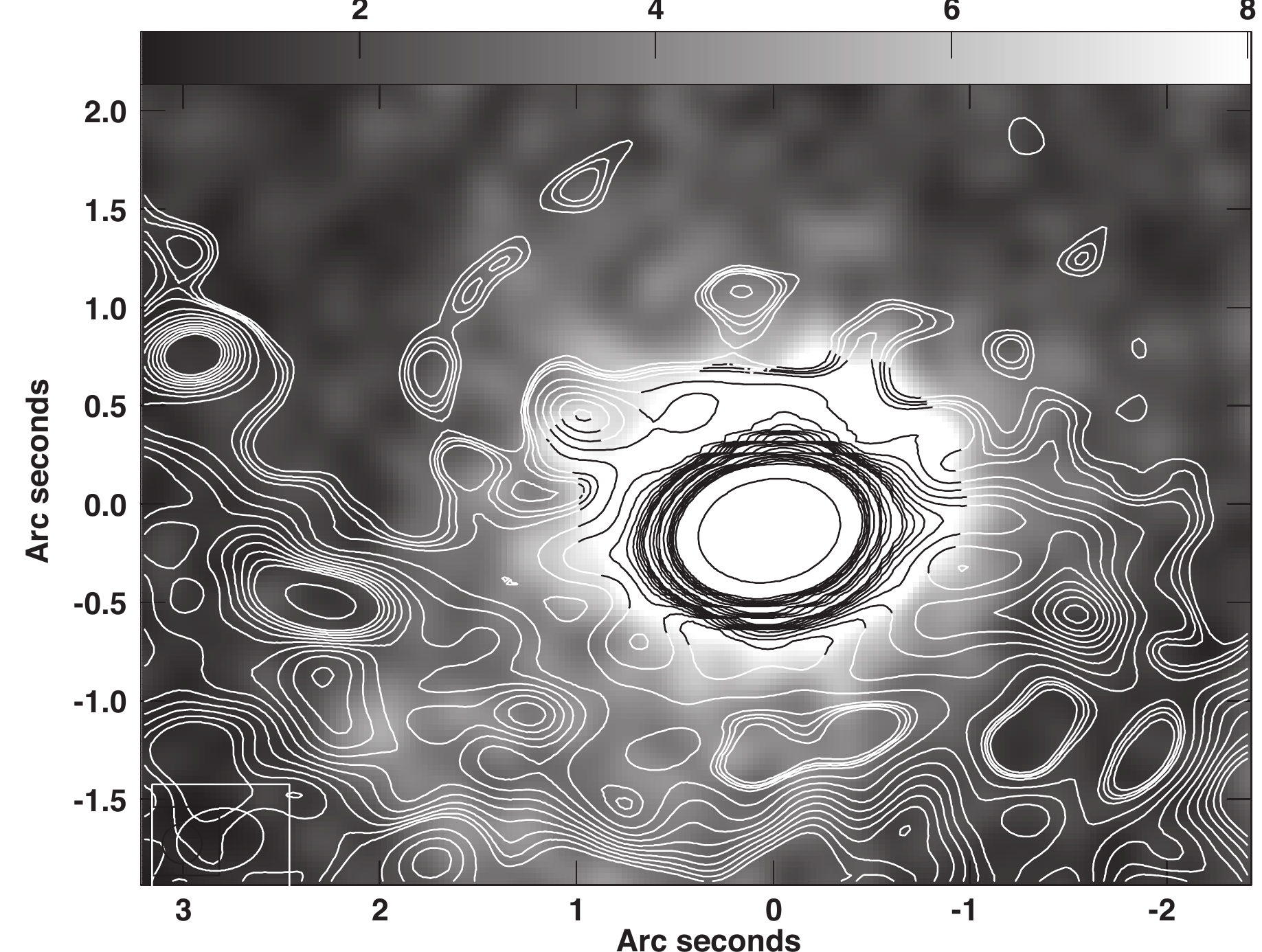}
\includegraphics[scale=0.6,angle=0]{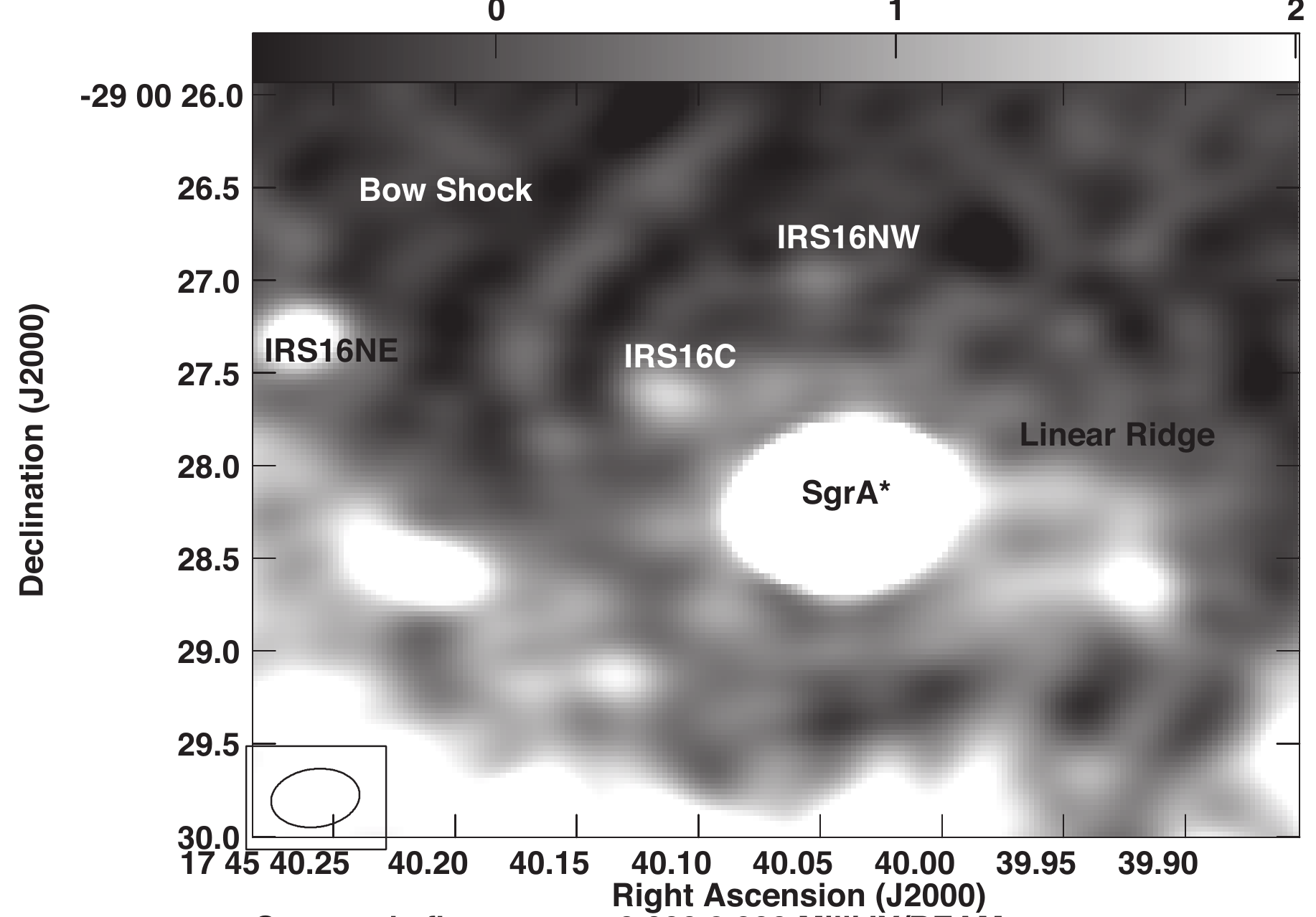}
\caption{
{\it (a)}
Contours of  226  GHz emission 
with levels set at 
(1, 2,...,10, 12, 15, 20, 25, 30, 35, 40, 50, 55, 60, 65, 70, 80, 
90, 110, 130)$\times$0.15 mJy beam$^{-1}$ are superimposed on an 1.5-7 keV X-ray image 
of Sgr A*
with the  spatial resolution of  $\sim0.5''$. 
(The grayscale range 0 to 8$\times10^3$ counts.) 
{\it (b)}
A grayscale image of the  mm emission from the inner 4$''\times6''$ of Sgr A* 
with  similar resolution to the 226 GHz image in (a). 
 Prominent stellar sources,
the bow shock  and the  linear features are labeled.  
(The grayscale range $-1\times10^{-3} - 3\times10^{-3}$ Jy beam$^{-1}$.)
}
\end{figure}

\begin{figure}
\center
\includegraphics[scale=0.5,angle=0]{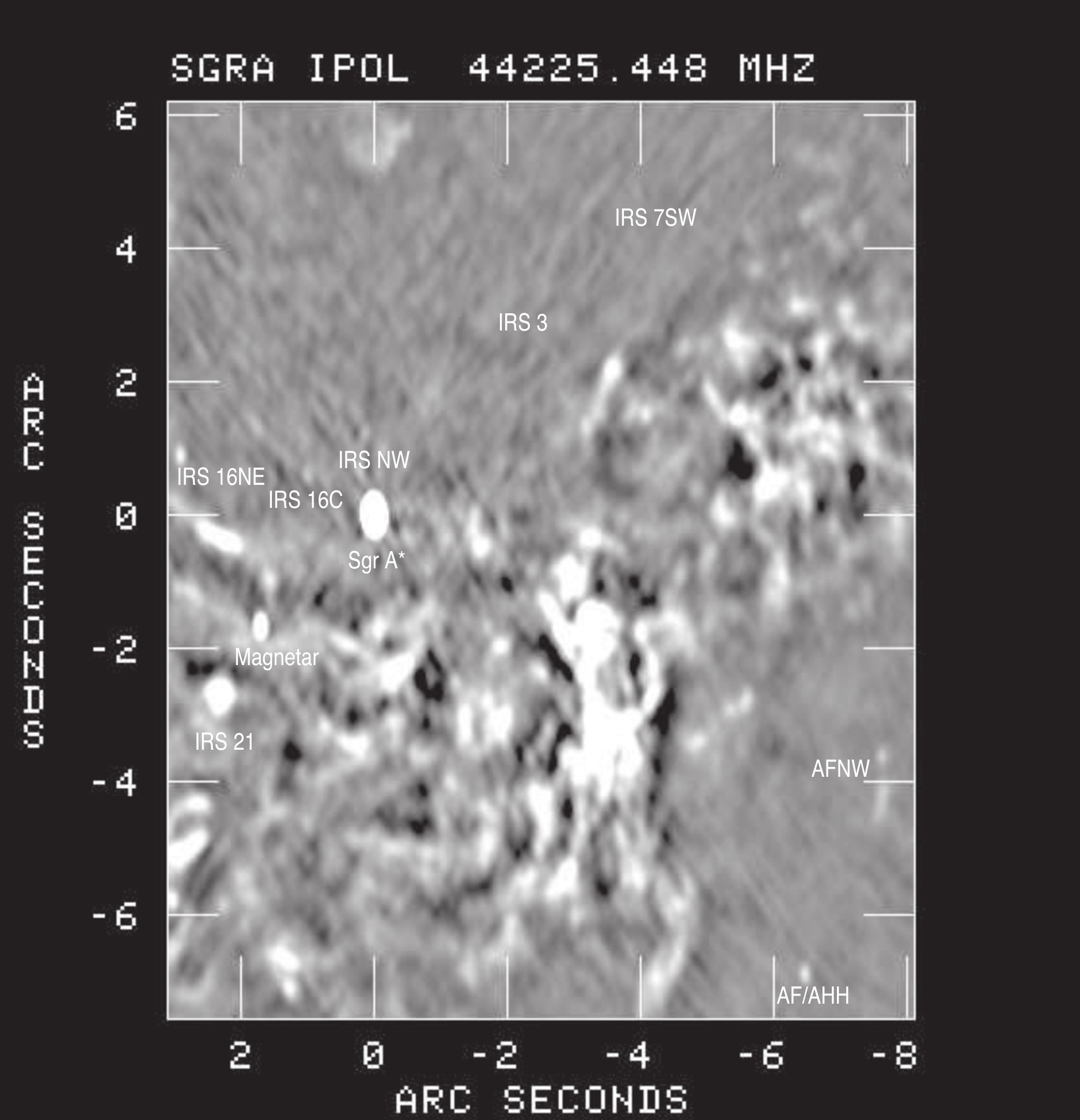}
\includegraphics[scale=0.5,angle=0]{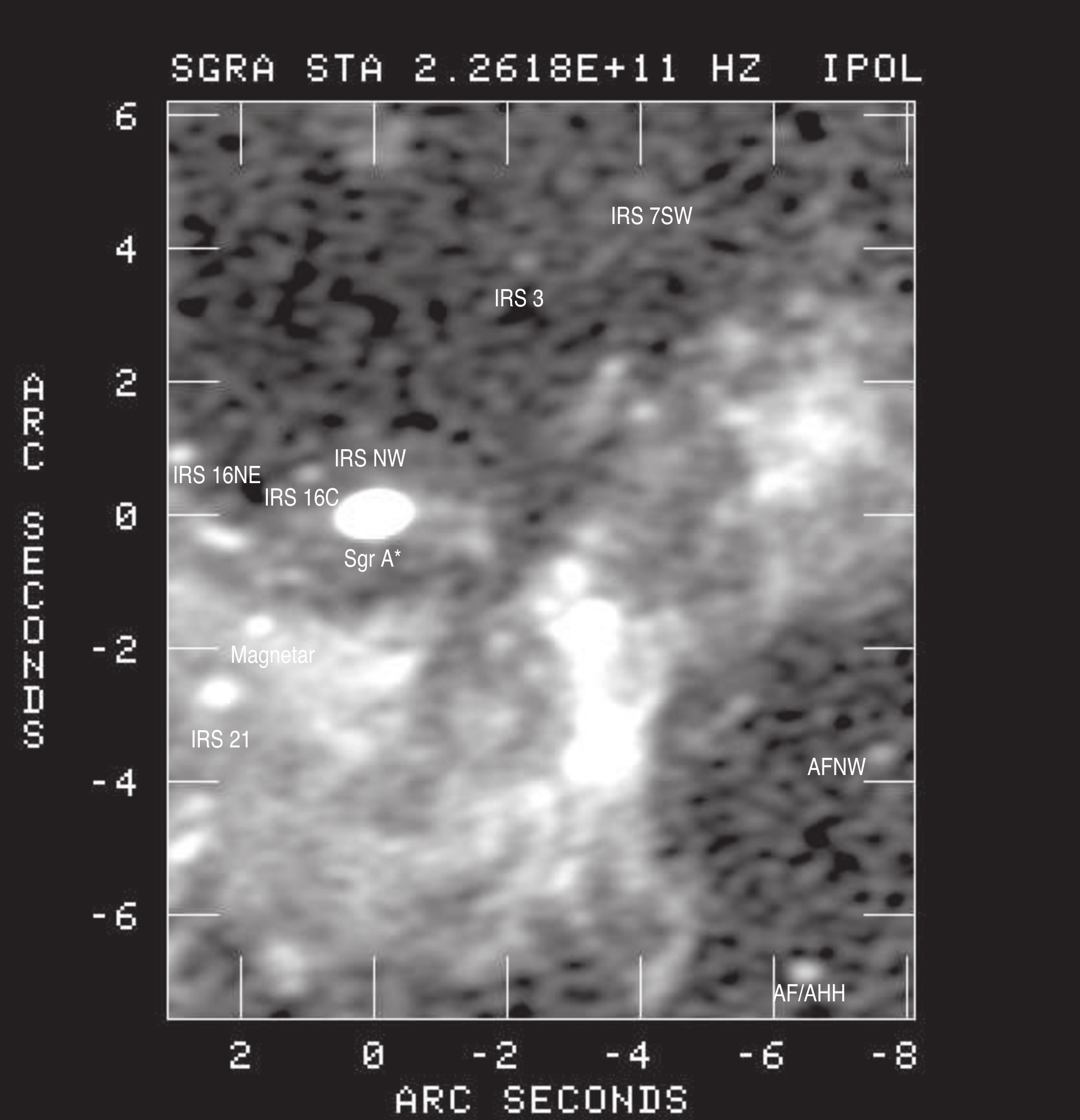}
\caption{
{\it (a) Top}
A 44  GHz image of the mini-spiral is constructed by limiting the {\it{uv}} range to greater than 
100 k$\lambda$ with  a spatial
resolution of $0.22''\times0.13''$ (PA=$3.8^\circ$) from the 2016, July 12 epoch.   
(The grayscale range $-5\times10^{-4} - 5\times10^{-4}$ Jy beam$^{-1}$.)
{\it (b) Bottom}
A 226  GHz image of the mini-spiral with similar resolution to that of Figure 1a. 
Labeled sources are compact with properties listed in Tables 1 and  2. 
(The grayscale range $-6.7\times10^{-4} - 1\times10^{-2}$ Jy beam$^{-1}$.)
}
\end{figure}


\begin{figure}
\center
\includegraphics[scale=0.4,angle=0]{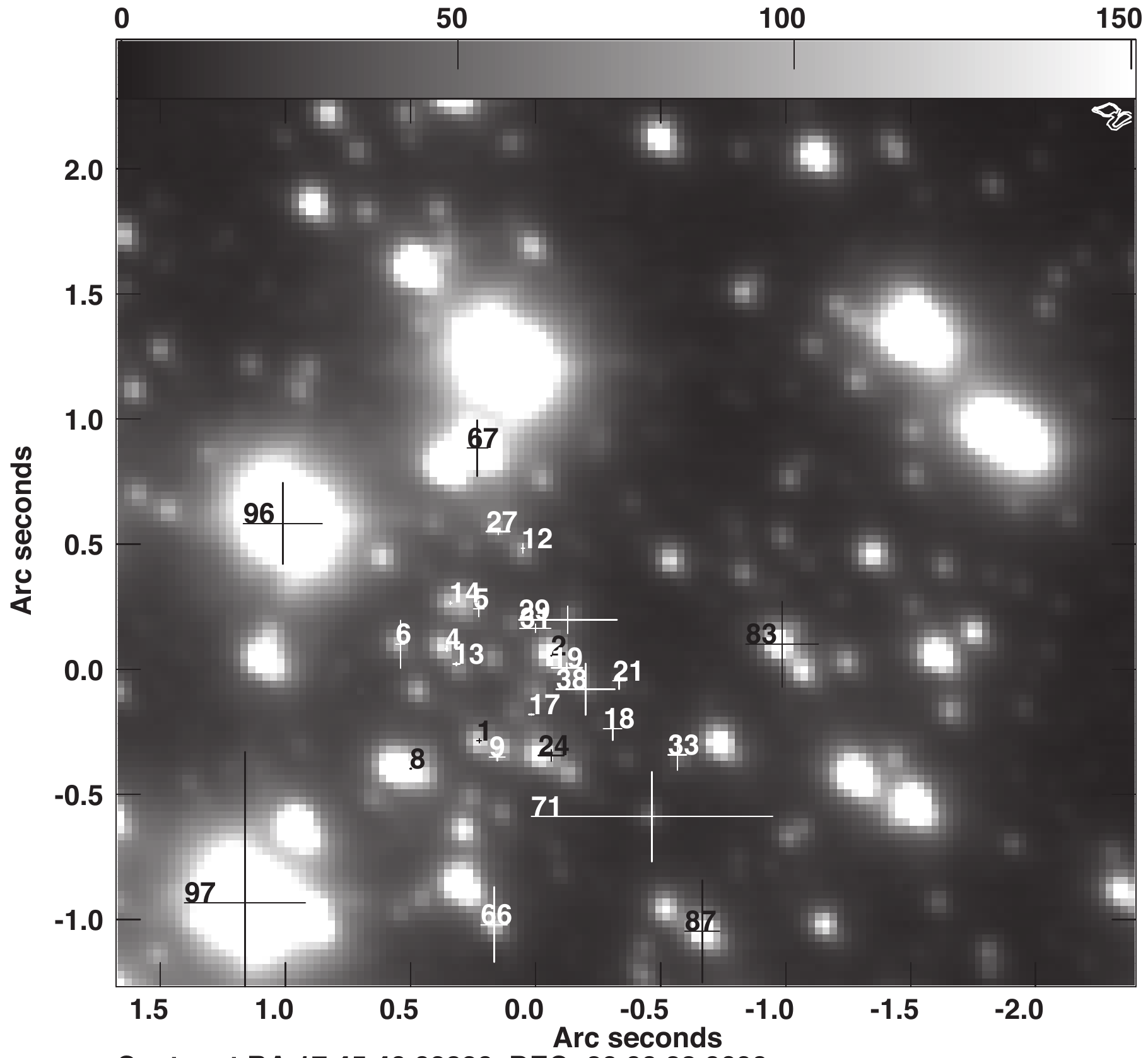}
\includegraphics[scale=0.4,angle=0]{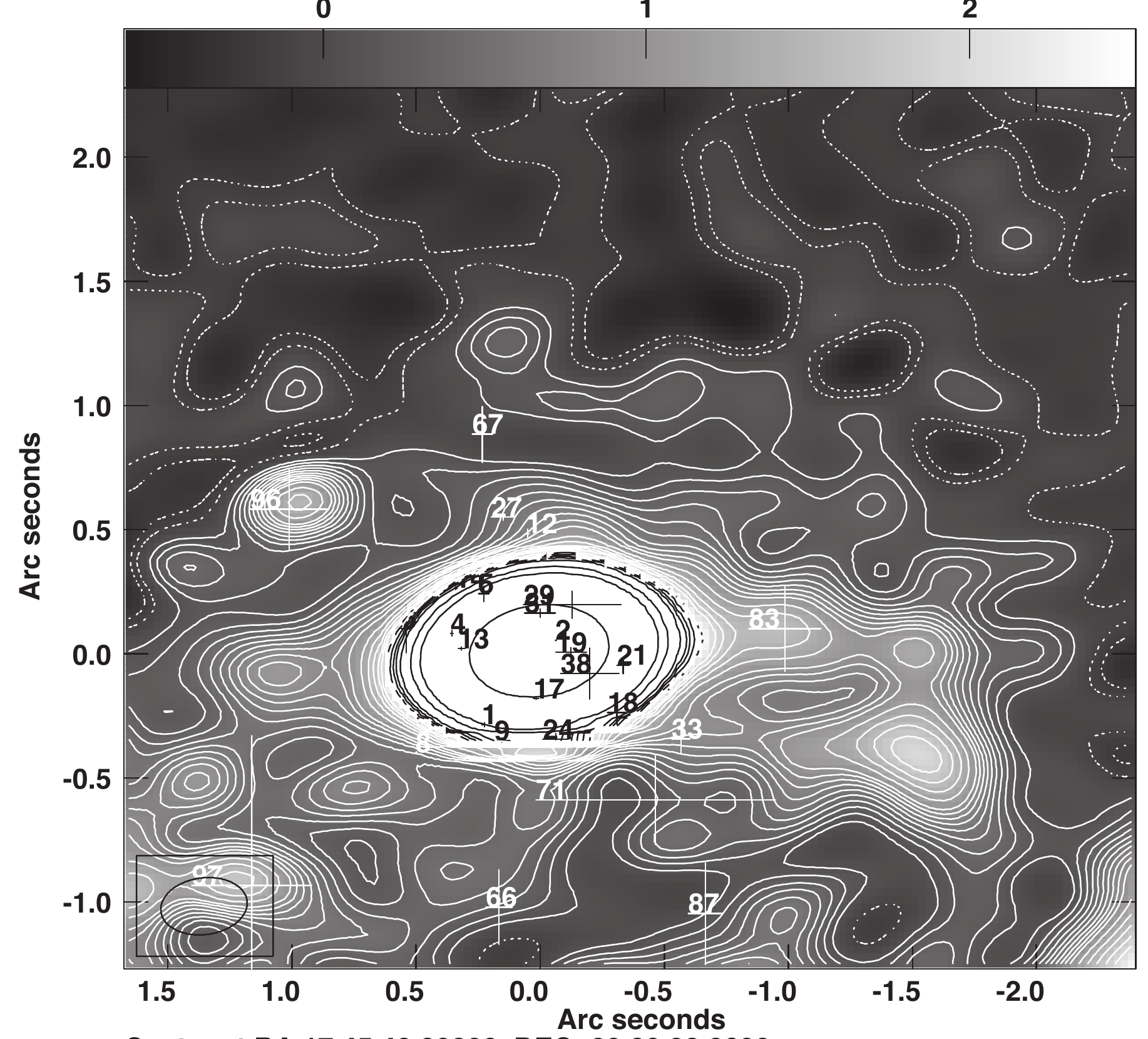}
\includegraphics[scale=0.4,angle=0]{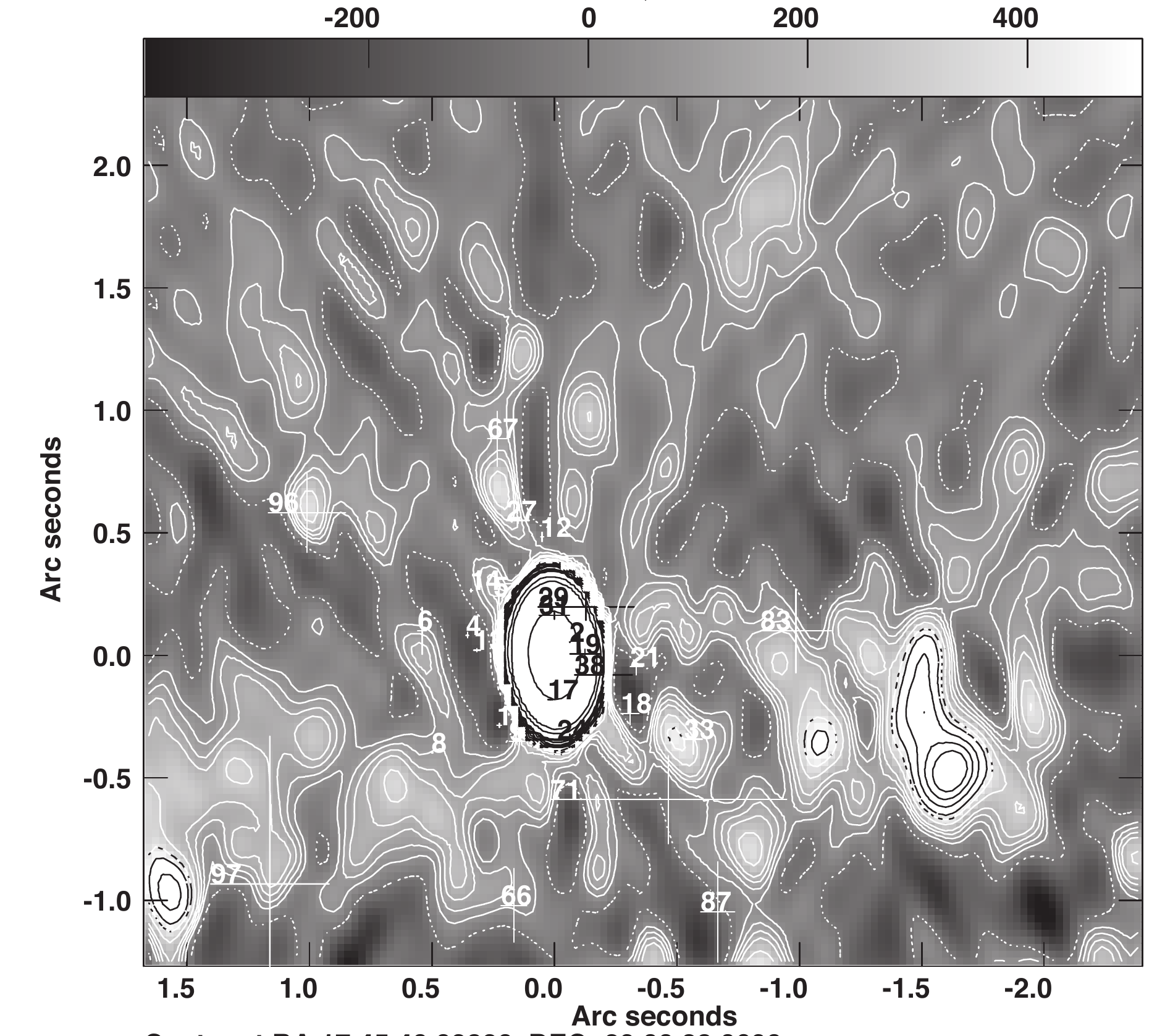}
\includegraphics[scale=0.4,angle=0]{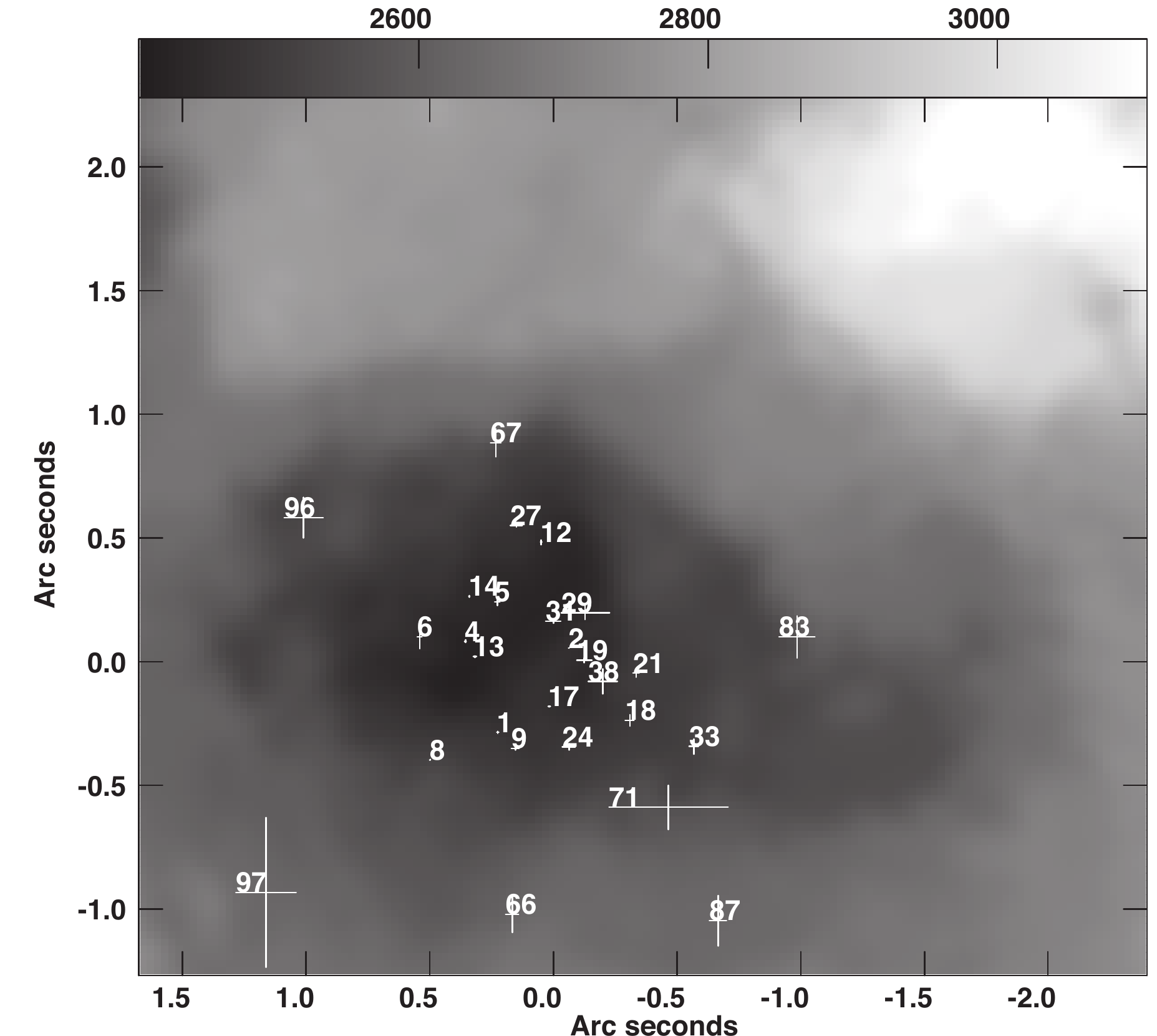}
\caption{
{\it (a) Top Left}
A grayscale image of the inner 3.5$''$ of Sgr A* at H band taken on 2016, July 12.
{\it (b) Top Right}
Grayscale contours of  226  GHz emission 
set at (-2, -1, 1, 2,...,10, 12,.., 20, 25, 30, 35, 40, 50, 100, 200, 5000)$\times$0.1 mJy beam$^{-1}$ 
 with a   spatial resolution of  0.35$''\times0.23''$  (PA=$-82.7^\circ$). 
{\it (c) Bottom Left} Grayscale contours of 44.2 GHz emission set at
(-2, -1, 1, 2,...,10, 12,..,20, 25, 30, 35, 40, 50, 100, 200, 5000)$\times$0.05 mJy beam$^{-1}$
 with a   spatial resolution of  0.22$''\times0.13''$  (PA=$3.8^\circ$), taken on 2016, July 12. 
{\it (d) Bottom Right}
An extinction map showing an elongated  dust cavity coincident with the S cluster (Scho\"odel et al. 2009). 
Stellar sources associated with the S cluster are labeled at their expected positions on 2016, July 12.
The extinction ranges between 2.4 and 3.14 magnitudes.
}
\end{figure}

\begin{figure}
\center
\includegraphics[scale=0.5,angle=0]{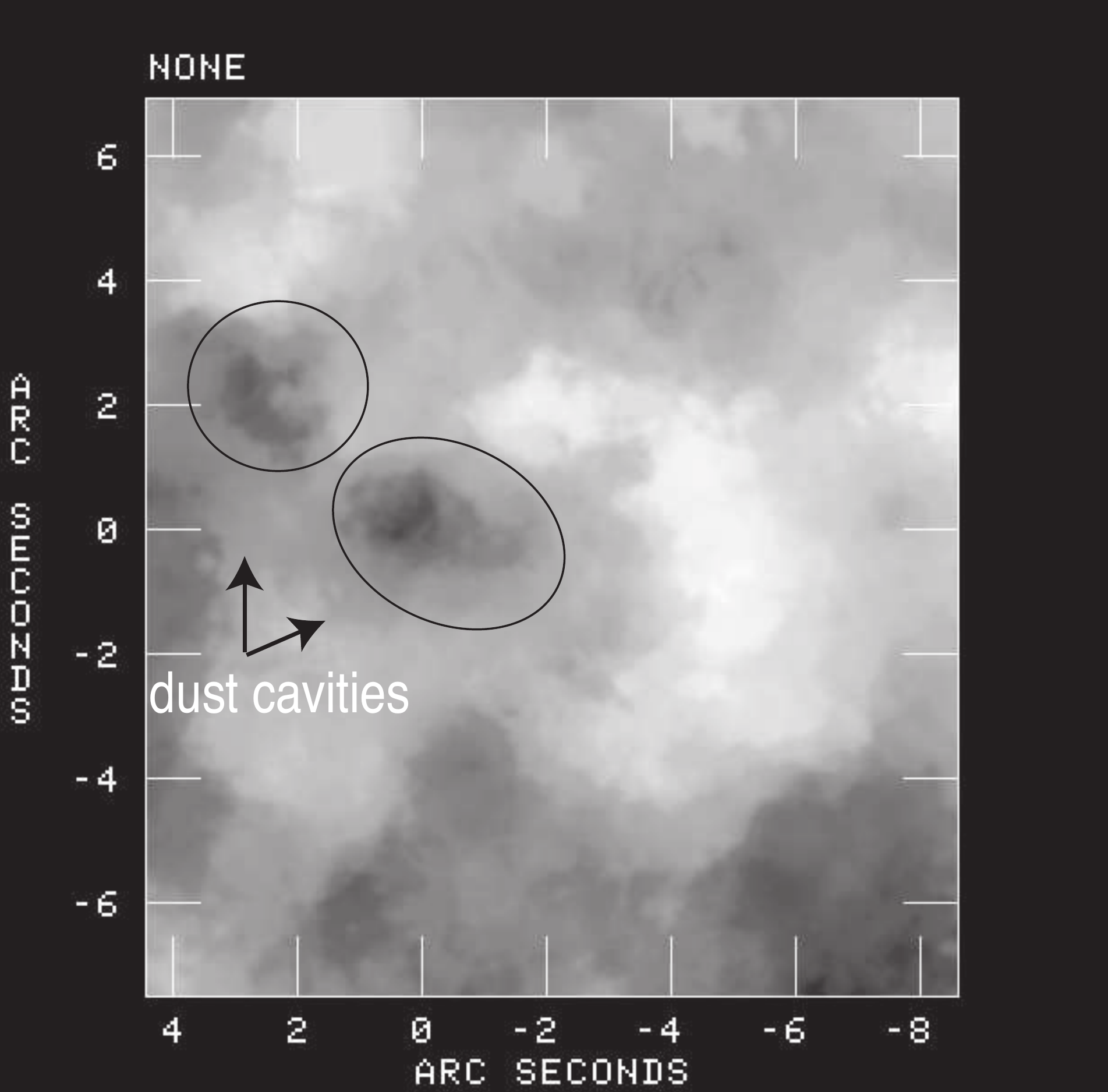}
\includegraphics[scale=0.5,angle=0]{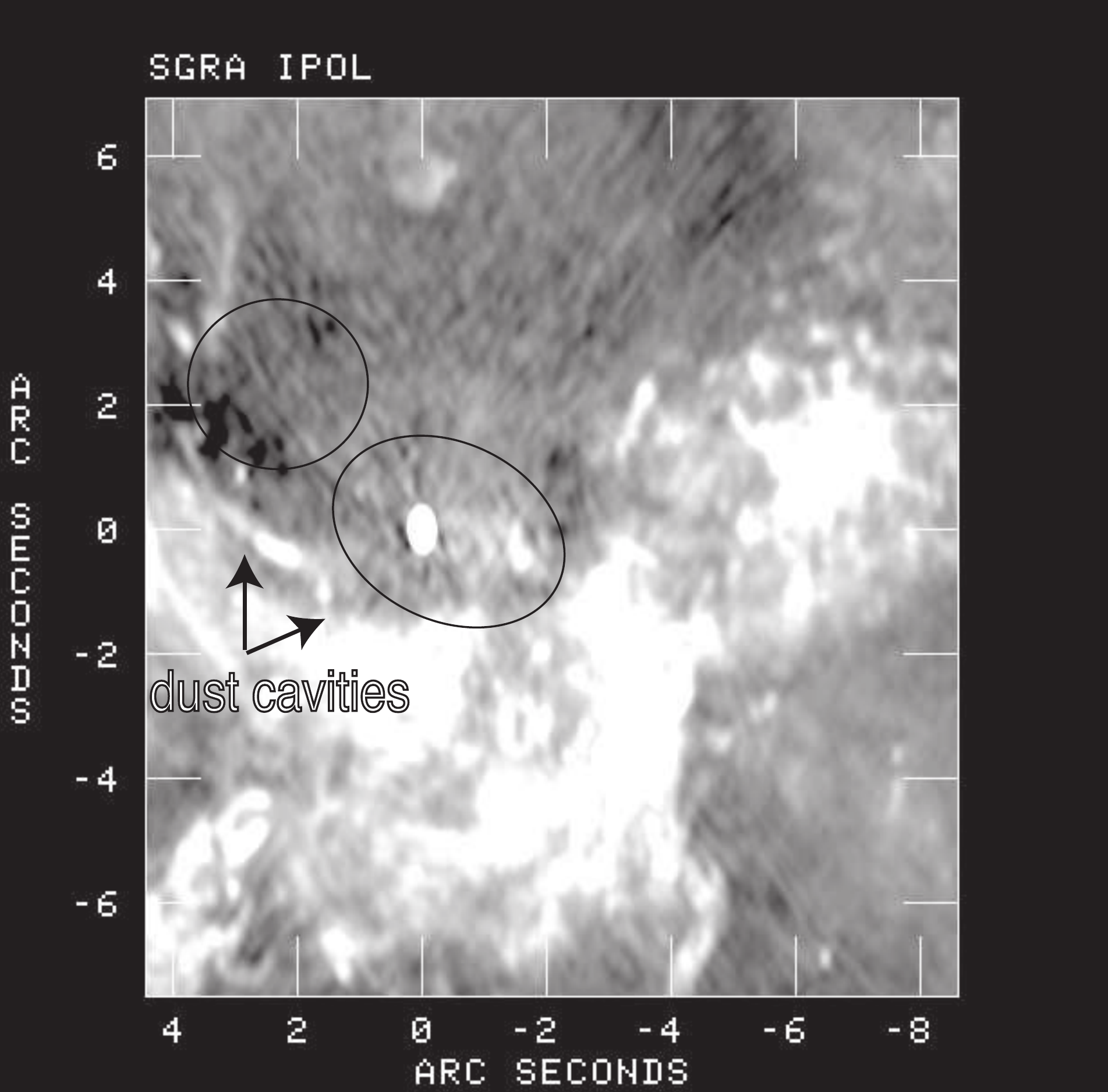}
\caption{
{\it (a) Top}
Similar to the region shown in Figure 4d except   the inner 13$''\times13''$ of Sgr A*. 
The extinction ranges between 2.4 and 3.20 magnitudes.
{\it (b) Bottom}
Similar to the region shown in  Figure 3a  except   that  {\it {uv}} data was not truncated, resulting 
a  spatial resolution of  0.24$''\times0.14''$  (PA=$3.3^\circ$). 
(The grayscale range $-3\times10^{-4} - 1\times10^{-3}$ Jy beam$^{-1}$.)
}
\end{figure}

\begin{figure}
\center
\includegraphics[scale=0.7,angle=0]{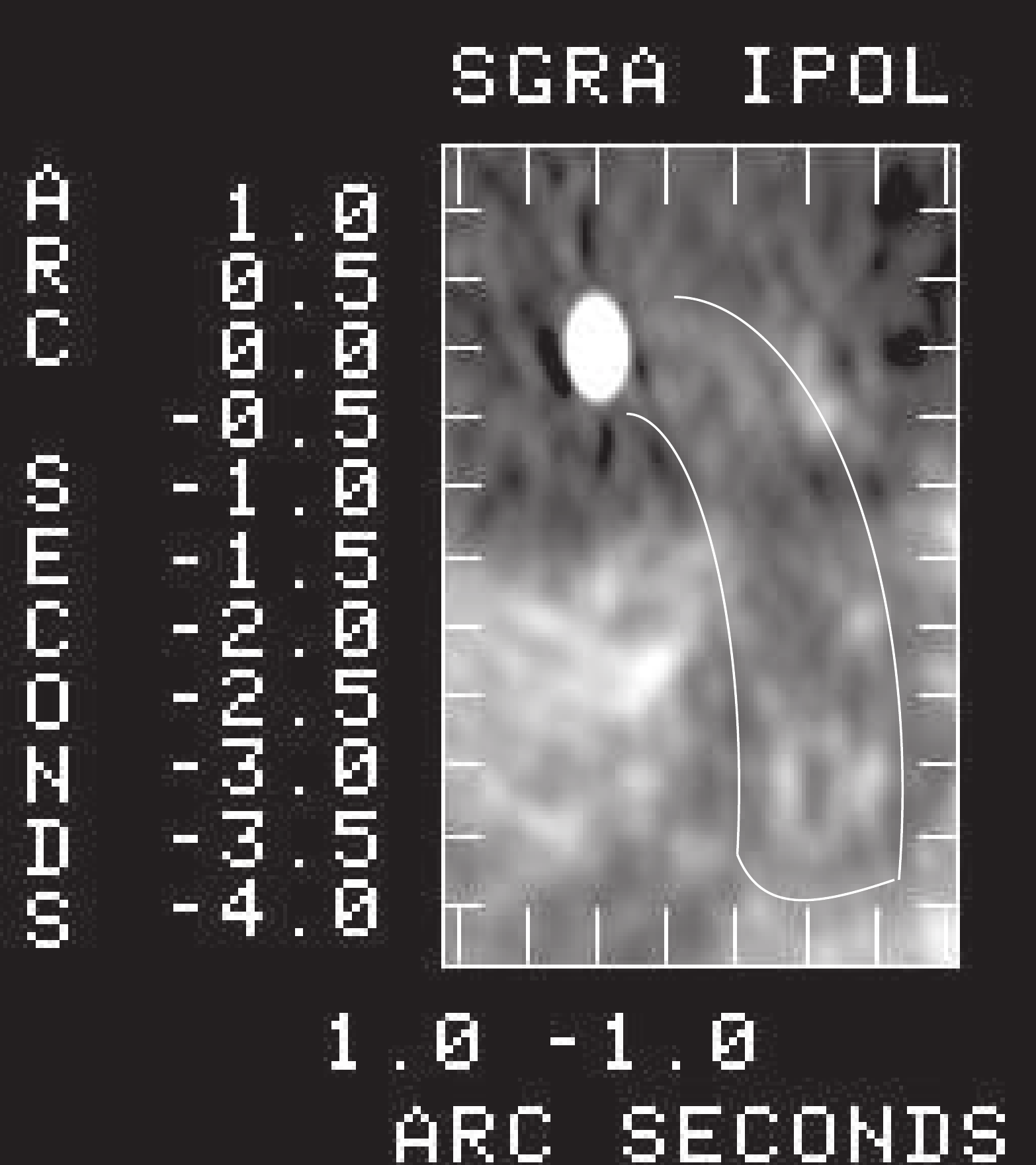}\\
\includegraphics[scale=0.45,angle=0]{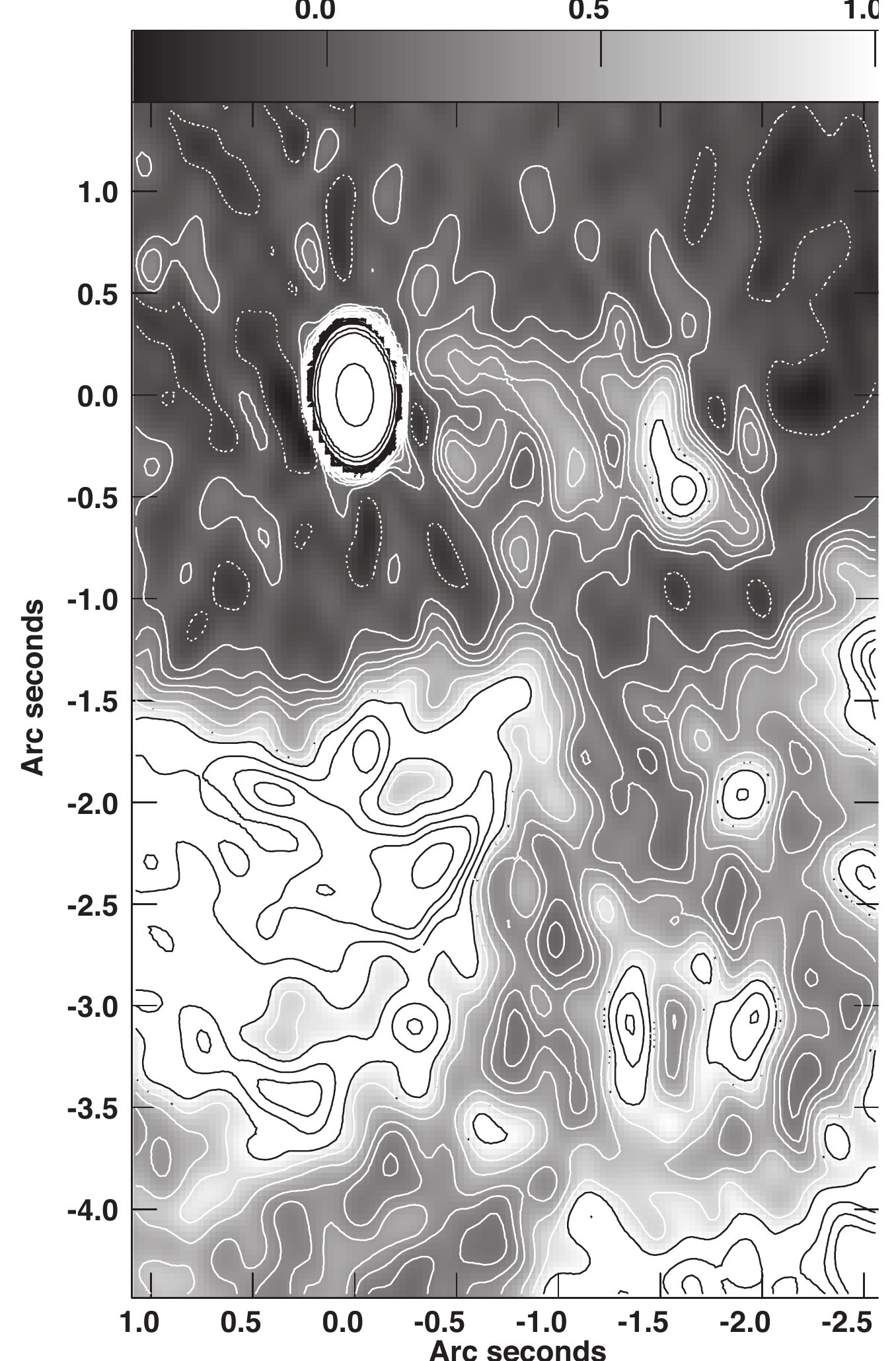}
\includegraphics[scale=0.45,angle=0]{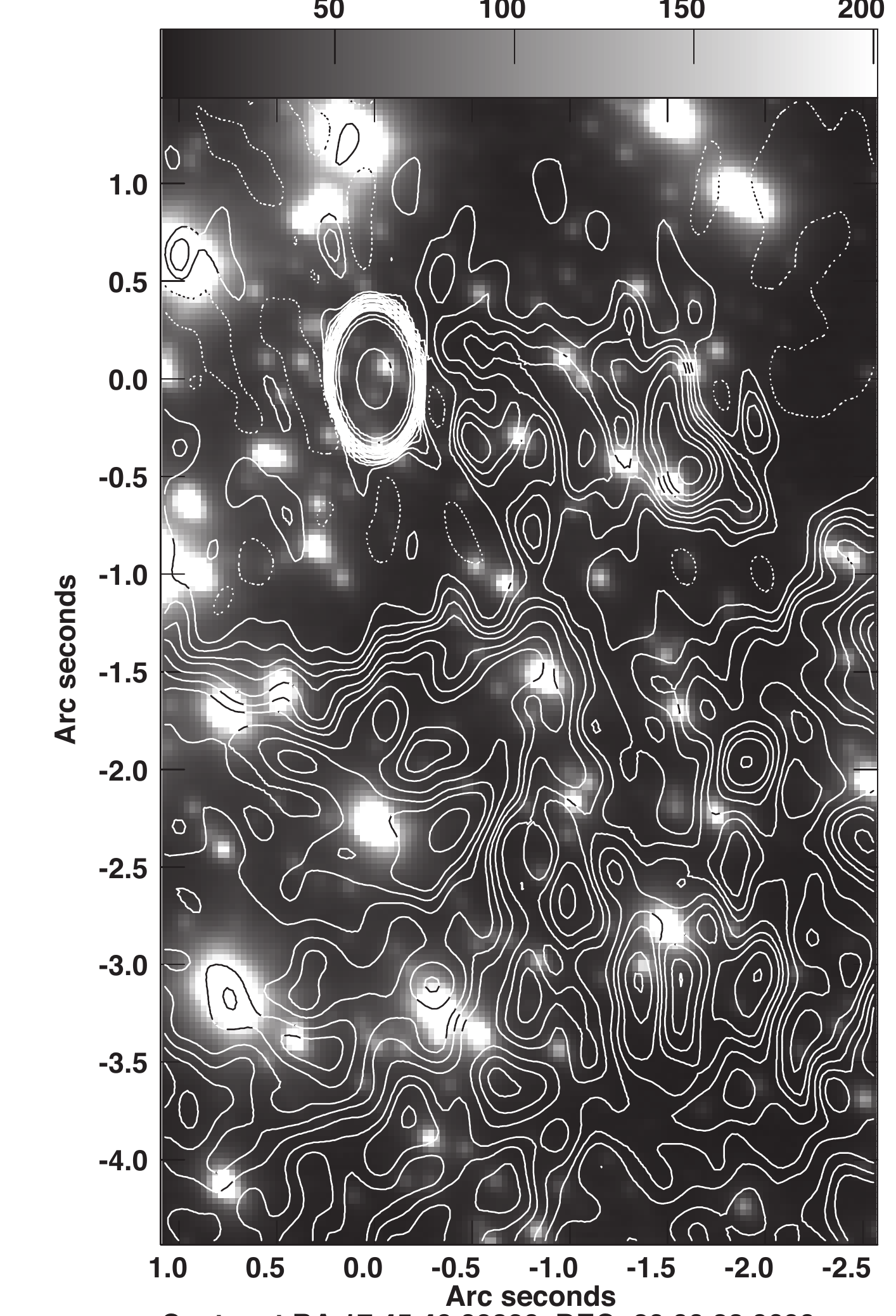}
\caption{
{\it (a) Top} 
Similar to Figure 5b  except   the inner  3.5$''\times6''$ of Sgr A*. 
{\it (b) Bottom Left} 
Similar to (a) except showing grayscale contours of emission set at
(-2, -1, 1, 2,...,10, 14, 18, 22, 30, 38, 46, 100, 200, 5000)$\times$0.1 mJy beam$^{-1}$. 
{\it (c) Bottom Right} 
Similar to (b) except that the 44 GHz contours are superimposed on a H-band image. 
}
\end{figure}

\begin{figure}
\center
\includegraphics[scale=0.5,angle=0]{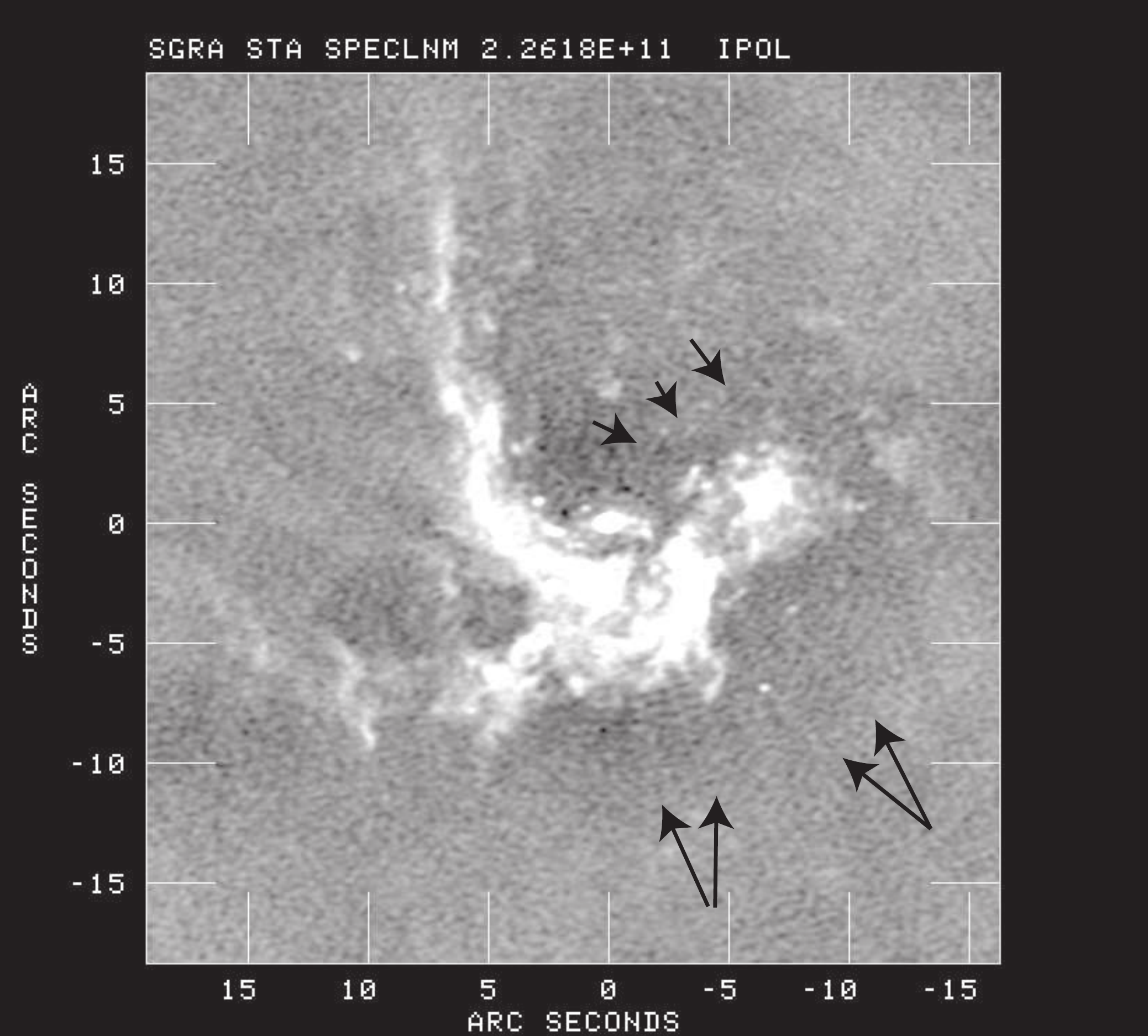}
\includegraphics[scale=0.5,angle=0]{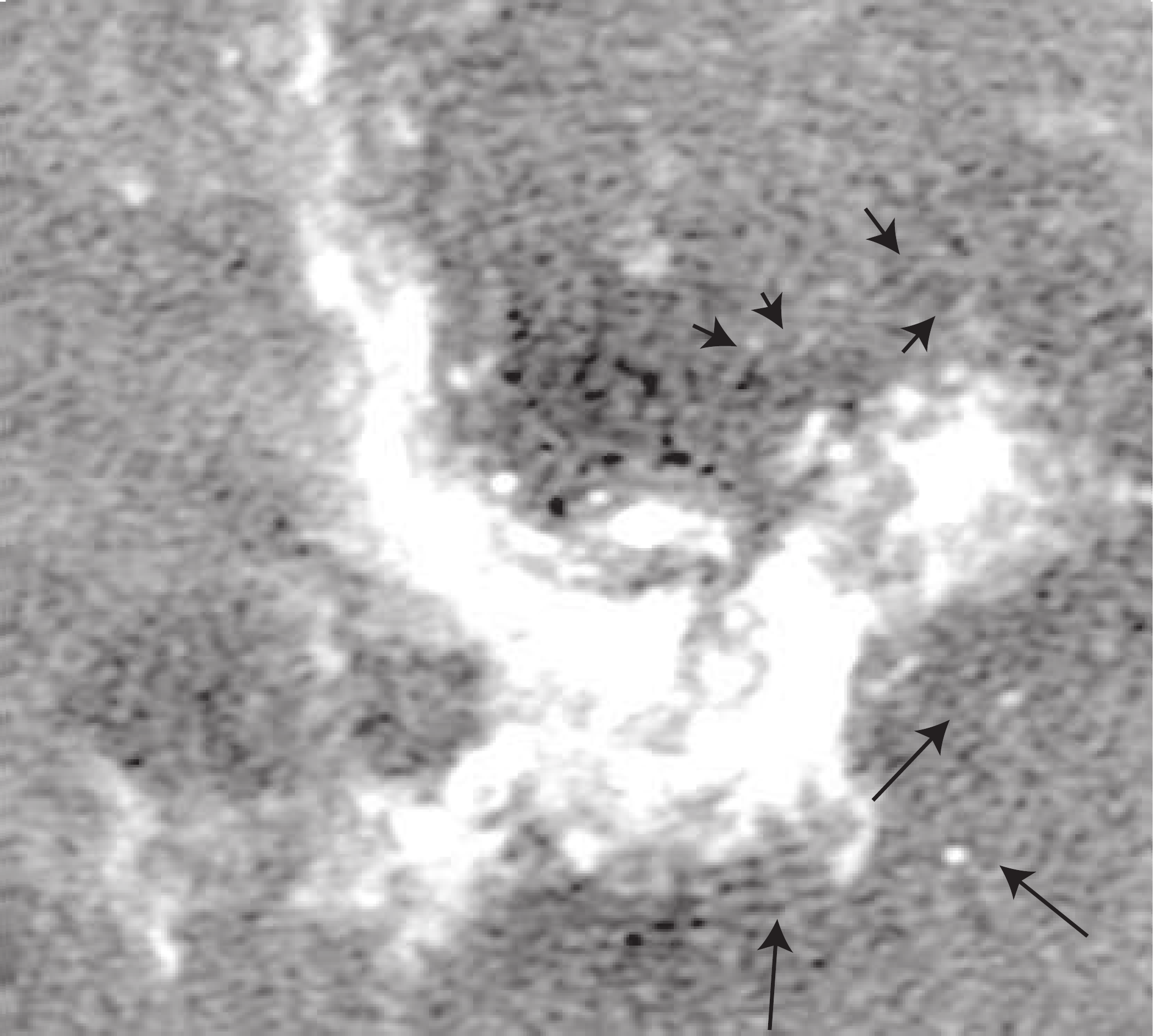}
\caption{
{\it (a) Top}
A Grayscale 226 GHz image of  the mini-spiral  with a
 spatial resolution  of  0.22$''\times0.13''$  (PA=$3.8^\circ$)
taken on 2016, July 18  with ALMA. 
{\it (b) Bottom}
The inner quarter of the 226 GHz image shown in (a) with a resolution of 0.36$''\times0.24''$ (PA$=-82^\circ.4$.
The arrows point to faint fibrils detected in radio and mm images.
}
\end{figure}

\begin{figure}
\center
\includegraphics[scale=0.6,angle=0]{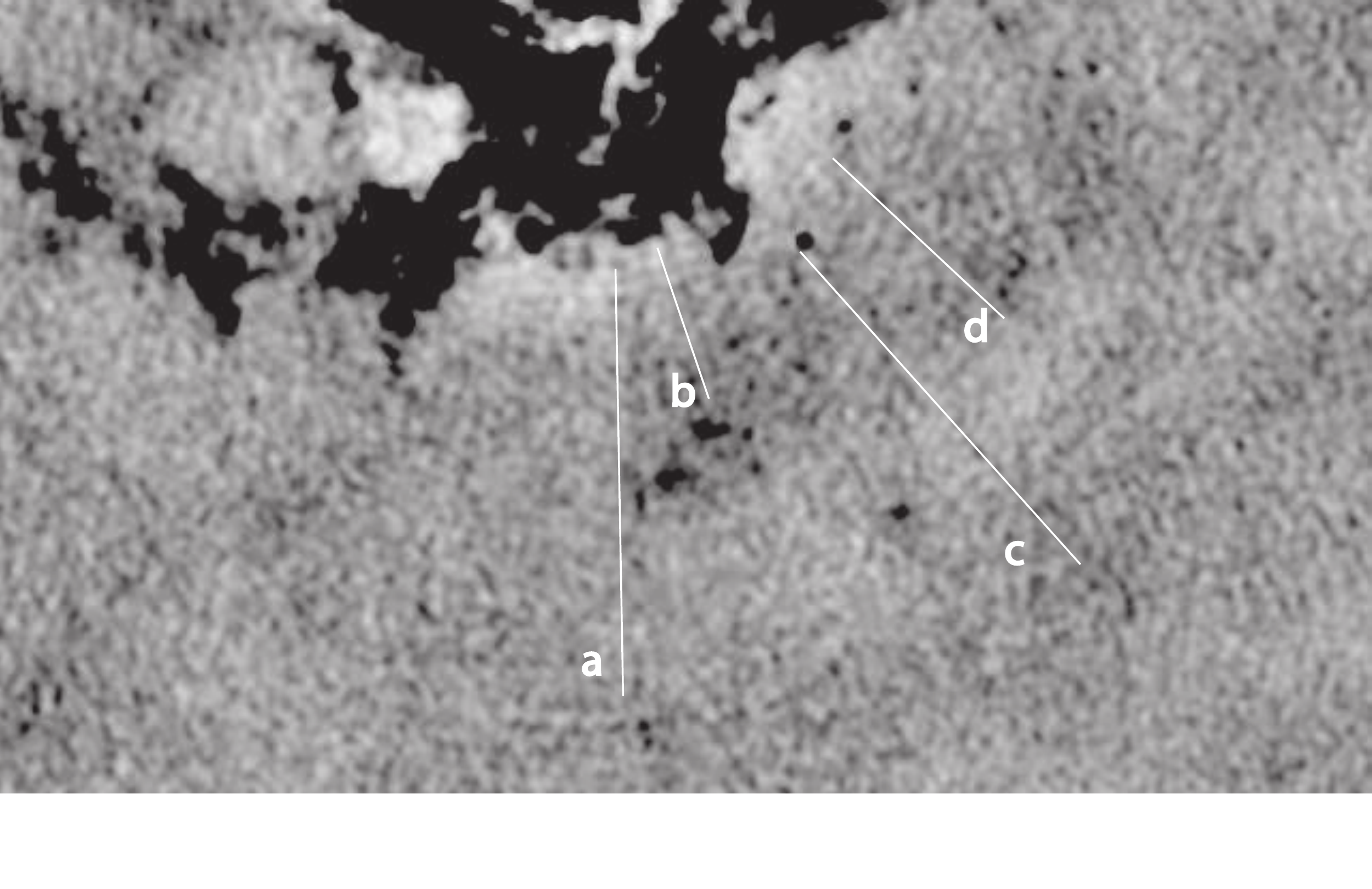}
\includegraphics[scale=0.4,angle=0]{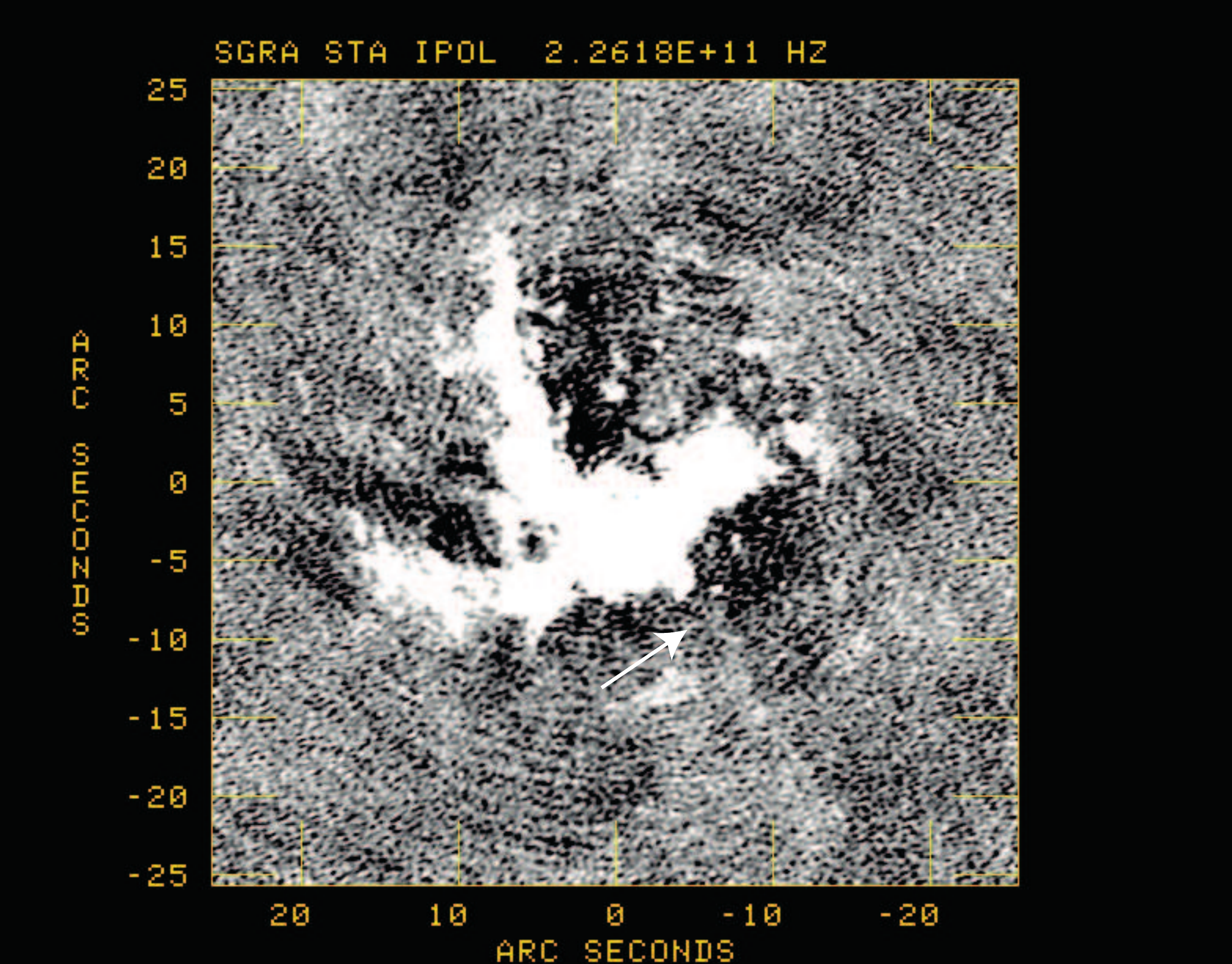}
\caption{
{\it (a) Top}
A 226 GHz image (in reverse color) with a resolution of 0.45$''\times0.45''$ from 
the 2016, July 18 data taken with ALMA. The drawn 
paralled lines point where the fibrils are located. 
{\it (b) Bottom} 
A 226 GHz image with  
a spatial resolution  of  0.44$''\times0.34''$  (PA=$-70.8^\circ$)
taken on 2016, July 12 with ALMA.  
(The grayscale range $-3\times10^{-4} - 5\times10^{-4}$ Jy beam$^{-1}$.)
}
\end{figure}

\begin{figure}
\center
\includegraphics[scale=0.4,angle=0]{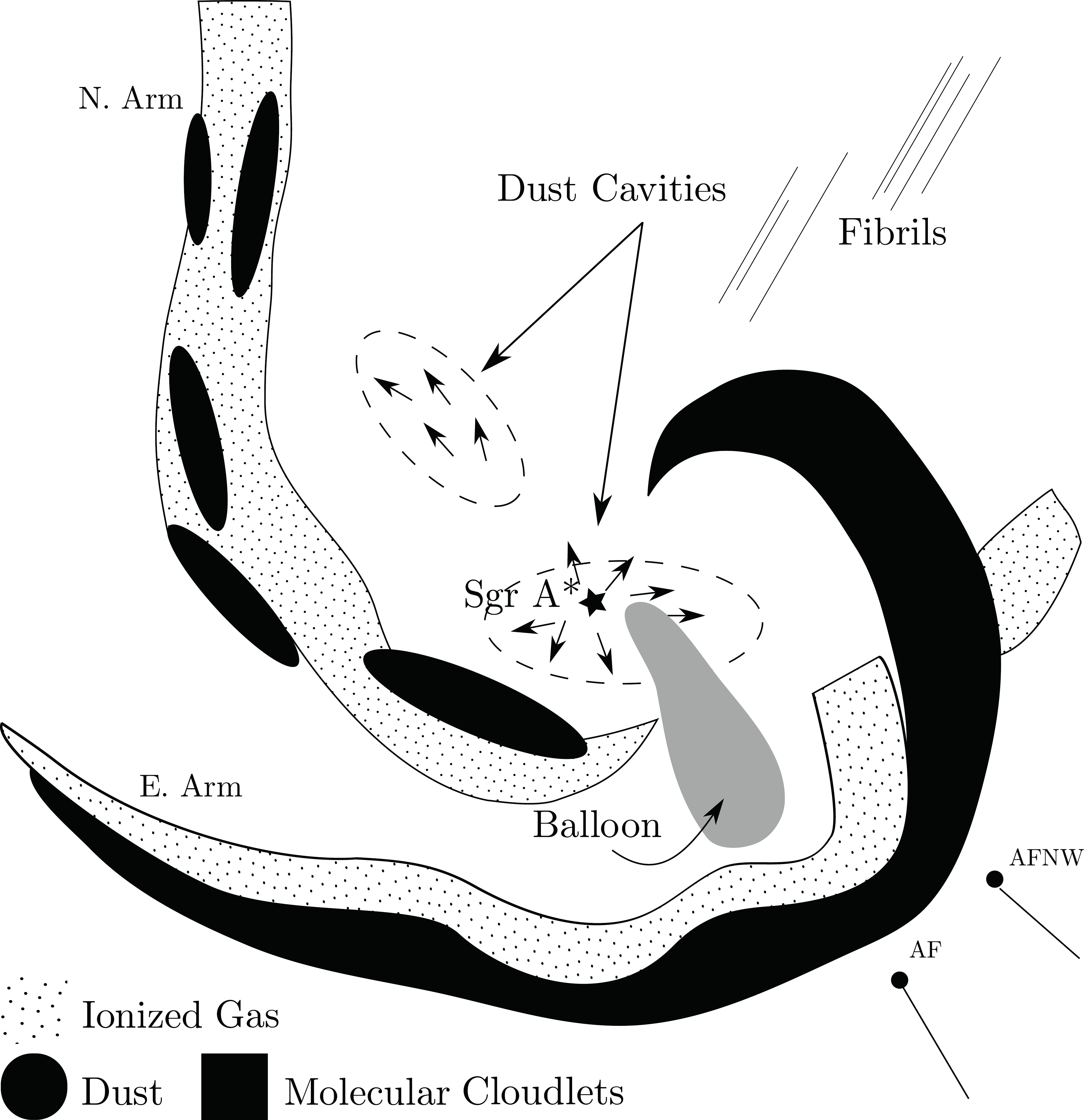}
\caption{A schematic picture of major features found at 226 GHz. Dark features 
point to regions where  extended dust and neutral gas is detected.  
}
\end{figure}

\bsp	
\label{lastpage}

\begin{landscape}
\begin{table}
\caption{Parameters of 2D Gaussian fits to 44 GHz Stellar Sources}
\begin{tabular}{llcccclcccl}
{ID} & {Alt Name} & {RA (J2000)} & {Dec (J2000)} & {Distance} & {Position}  
& {$ \theta_{a} \times \theta_{b} $ (PA)} & {Peak Intensity} & 
{Spectral Index} & {Integrated Flux} \\
& &  &  & from Sgr A* & Accuracy & &  &  &  \\
& & ($17^{\rm h}45^{\rm m}$) & ($-29^{\circ}00^{\prime}$) & (\sl{arcsec}) & (\sl{mas}) &
 \sl{mas} $ \times $ \sl{mas} (\sl{deg}) & \sl{(mJy beam$^{-1}$)} & ($\alpha$) &  \sl{(mJy)}\\
1 & Sgr A* & 40.0383 & 28.0690 & 0.00 & 0.00 & 3 $ \times $ 0 (31)
 & 1536.400 $ \pm $ 0.052 & 0.56 $ \pm $ 0.00004 & 1536.900 $ \pm $ 0.090  \\
2 & IRS 16C & 40.1134 & 27.4370 & 1.17 & 12.24 & 193 $ \times $ 155 (177)
 & 0.502 $ \pm $ 0.049 & 0.67$\pm0.29$ & 1.044 $ \pm $ 0.143  \\
3 & IRS 16NW & 40.0487 & 26.8459 & 1.23 & 16.41 & 109 $ \times $ 0 (149)
 & 0.317 $ \pm $ 0.052 & 0.28$\pm0.81$ & 0.293 $ \pm $ 0.085  \\
4 & Magnetar & 40.1685 & 29.7421 & 2.39 & 0.88 & 74 $ \times $ 0 (157)
 & 5.787 $ \pm $ 0.052 & -0.21$\pm0.10$ & 5.866 $ \pm $ 0.091  \\
5 & IRS 16NE & 40.2608 & 27.1745 & 3.05 & 5.48 & 100 $ \times $ 0 (13)
 & 0.977 $ \pm $ 0.052 & 0.79$\pm0.12$ & 1.056 $ \pm $ 0.095  \\
6 & IRS21 & 40.2154 & 30.7693 & 3.56 & 1.63 & 258 $ \times $ 236 (98)
 & 4.049 $ \pm $ 0.048 & 0.23$\pm0.07$ & 13.309 $ \pm $ 0.199  \\
7 & IRS 3 & 39.8614 & 24.2147 & 4.50 & 50.24 & 181 $ \times $ 0 (139)
 & 0.109 $ \pm $ 0.052 & 1.14$\pm0.66$ &  0.121 $ \pm $ 0.096  \\
8 & IRS 7SW & 39.7383 & 23.2063 & 6.26 & 63.47 & 250 $ \times $ 0 (16)
 & 0.111 $ \pm $ 0.051 & 1.12$\pm0.63$ & 0.147 $ \pm $ 0.107  \\
9 & AFNW & 39.4558 & 31.6992 & 8.46 & 12.69 & 95 $ \times $ 43 (151)
 & 0.411 $ \pm $ 0.052 & 0.73$\pm0.31$ &  0.488 $ \pm $ 0.100  \\
10 & AF/AHH & 39.5446 & 34.9672 & 9.46 & 6.60 & ---
 & 0.735 $ \pm $ 0.052 & 0.84$\pm0.15$ & 0.717 $ \pm $ 0.088  \\
\end{tabular}
\end{table}
\end{landscape}

\begin{landscape}
\begin{table}
\caption{Parameters of 2D Gaussian fits to 226 GHz Stellar Sources}
\begin{tabular}{llcccclccl}
{ID} & {Alt Name} & {RA (J2000)} & {Dec (J2000)} & {Dist. from Sgr A*} & 
 {Pos. Accuracy} & {$ \theta_{a}\times\theta_{b}$(PA)} & {Peak Intensity} & 
{Integrated Flux}\\
1 & Sgr A* & 40.0386 & 28.0580 & 0.00 & 0.02 & 40 $ \times $ 31 (100)
 & 2970.000 $ \pm $ 0.682 & 3017.500 $ \pm $ 1.190 & \\
2 & IRS 16C & 40.1115 & 27.4709 & 1.12 & 43.60 & 29 $ \times $ 0 (143)
 & 1.490 $ \pm $ 0.682 & 1.396 $ \pm $ 1.130 & \\
3 & IRS 16NW$^2$ & 40.0484 & 26.8348 & 1.23 & 187.43 & 243 $ \times $ 116 (174)
 & 0.501 $ \pm $ 0.660 & 0.767 $ \pm $ 1.540 & \\
4 & J1745-29 & 40.1700 & 29.7533 & 2.41 & 18.41 & 128 $ \times $ 67 (156)
 & 4.078 $ \pm $ 0.679 & 4.699 $ \pm $ 1.290 & \\
5 & IRS 16NE & 40.2623 & 27.1747 & 3.07 & 20.39 & 97 $ \times $ 81 (150)
 & 3.535 $ \pm $ 0.682 & 3.921 $ \pm $ 1.270 & \\
6 & IRS 21 & 40.2160 & 30.7499 & 3.56 & 14.20 & 228 $ \times $ 185 (104)
 & 5.877 $ \pm $ 0.660 & 9.044 $ \pm $ 1.540 & \\
7 & IRS 3$^2$ & 39.8644 & 24.2573 & 4.43 & 101.29 & 93 $ \times $ 0 (102)
 & 0.701 $ \pm $ 0.682 & 0.641 $ \pm $ 1.110 & \\
8 & IRS 7SW\footnote{To improve the S/N ratio, combined data from
2016, July 13 and 19 are used to get flux densities
with a resolution 0.37$''\times0.26''$ (PA=-79$^\circ$.81)} & 39.7326 & 23.1742 & 6.32 & 129.46 & 408 $ \times $ 223 (77)
 & 0.696 $ \pm $ 0.643 & 1.521 $ \pm $ 1.940 & \\
9 & AFNW & 39.4577 & 31.6806 & 8.44 & 58.73 & 240 $ \times $ 46 (130)
 & 1.356 $ \pm $ 0.670 & 1.771 $ \pm $ 1.390 & \\
10 & AF/AHH & 39.5436 & 34.9422 & 9.46 & 25.10 & 166 $ \times $ 38 (52)
 & 2.910 $ \pm $ 0.677 & 3.449 $ \pm $ 1.320 & \\
\end{tabular}
\end{table}
\end{landscape}


\begin{landscape}
\begin{table}
\caption{Predicted Positions of Young  Stars in the S Cluster from Sgr A* at the 2016.54 Epoch}
\begin{tabular}{lccccccccc}
{Source Name} & RA (offset)  & Dec (offset) & {($\sigma_X$)} & {($\sigma_Y$)}\\
                                &  \sl{arcsec} & \sl{arcsec} & \sl{arcsec} & \sl{arcsec}\\
S1  &  0.2244  &  -0.2862  &   0.0161 &   0.0187 \\      
S2  & -0.0632  &  0.0536   &  0.0012  &   0.0038 \\      
S4  &  0.3550  &  0.0826   &  0.0226  &   0.0103 \\       
S5  &  0.2259  &  0.2417   &  0.0622  &   0.0429 \\      
S6  &  0.5399  &  0.0996   &  0.1895  &   0.0350 \\      
S8  &  0.4982  & -0.3972   &  0.0048  &   0.0063  \\     
S9  &  0.1531  & -0.3513   &  0.0272  &   0.0623  \\   
S12 &  0.0495  &  0.4826   &  0.0385  &   0.0125  \\     
S13 &  0.3160  &  0.0198   &  0.0128  &   0.0273 \\      
S14 &  0.3391  &  0.2641   &  0.0132  &   0.0103  \\     
S17 &  0.0161  & -0.1812   &  0.0040  &   0.0205  \\     
S18 & -0.3088  & -0.2381   &  0.0932  &   0.0719  \\     
S19 & -0.1253  &  0.0057   &  0.0381  &   0.1234  \\     
S21 & -0.3348  & -0.0473   &  0.0644  &   0.0498  \\     
S24 & -0.0634  & -0.3448   &  0.0472  &   0.1037 \\      
S27 &  0.1491  &  0.5496   &  0.0256  &   0.0944 \\      
S29 & -0.1297  &  0.1967   &  0.1094  &   0.3917 \\      
S31 & -0.0000  &  0.1630   &  0.0340  &   0.1237 \\      
S33 & -0.5679  & -0.3434   &  0.1219  &   0.0737  \\     
S38 & -0.1996  & -0.0801   &  0.2040  &   0.2352 \\      
S66 &  0.1655  & -1.0205   &  0.3000  &   0.1063 \\      
S67 &  0.2320  &  0.8833   &  0.2265  &   0.0823 \\      
S71 & -0.4660  & -0.5897   &  0.3592  &   0.9643 \\      
S83 & -0.9859  &  0.0999   &  0.3428  &   0.2893  \\     
S87 & -0.6667  & -1.0474   &  0.4086  &   0.1390 \\      
S96 &  1.0095  &  0.5818   &  0.3255  &   0.3163 \\      
S97 &  1.1609  & -0.9340   &  1.2096  &   0.4843 \\      
\end{tabular}
\end{table}
\end{landscape}
\end{document}